\documentclass{article}

\usepackage{microtype}
\usepackage{graphicx}
\usepackage{subcaption}
\usepackage{booktabs}

\usepackage{hyperref}

\usepackage[accepted]{icml2026}

\usepackage{amsmath}
\usepackage{amssymb}
\usepackage{mathtools}
\usepackage{amsthm}

\usepackage[capitalize,noabbrev]{cleveref}

\theoremstyle{plain}

\theoremstyle{definition}

\theoremstyle{remark}

\usepackage{relsize}
\usepackage{graphicx}
\usepackage[table]{xcolor}

\usepackage{ulem}
\robustify{\raisebox}
\usepackage{pifont}
\DeclareRobustCommand{\circledtwo}{\raisebox{-0.16ex}{\ding{183}}}
\DeclareRobustCommand{\circledone}{\raisebox{-0.16ex}{\ding{182}}}
\usepackage{soul}
\usepackage{bm}
\usepackage{booktabs}
\usepackage{multirow}
\usepackage{makecell}
\usepackage{textgreek}

\icmltitlerunning{New Wide-Net-Casting Jailbreak Attacks Risk Large Models}

\begin{document}

\twocolumn[
  \icmltitle{New Wide-Net-Casting Jailbreak Attacks Risk Large Models}

  \icmlsetsymbol{equal}{*}

  \begin{icmlauthorlist}
    \icmlauthor{Qiuchi Xiang}{yyy}
    \icmlauthor{Haoxuan Qu}{yyy}
    \icmlauthor{Hossein Rahmani}{yyy}
    \icmlauthor{Jun Liu}{yyy}
  \end{icmlauthorlist}

  \icmlaffiliation{yyy}{Lancaster University}

  \icmlcorrespondingauthor{Jun Liu}{j.liu81@lancaster.ac.uk}

  \icmlkeywords{Machine Learning, ICML}

  \vskip 0.3in
]

\printAffiliationsAndNotice{}

\begin{abstract}
Jailbreak attacks on large models have drawn growing attention due to their close ties to societal safety. This work identifies a practical yet unexplored jailbreak scenario, the wide-net-casting scenario, where an adversary can query a group of large models instead of a single one to elicit harmful outputs. Our analysis reveals substantial yet previously overlooked safety risks under this scenario. 
As a key part of our analysis, we further develop a novel jailbreak method tailored to the wide-net-casting scenario.
With this tailored method, the jailbreak success rate can even reach 100\% in some experiments when targeting the large models without additional safeguards, exposing wide-net-casting as a distinct, high-risk scenario that warrants attention in future evaluation and defense research. Code is available \href{https://zzlz233.github.io/Wide-net-casting/}{here}.
\textcolor{red}{Warning: This paper contains potentially harmful example text.}
\end{abstract}

\section{Introduction}
\label{sec: intro}

Large language models (LLMs) and multimodal large language models (MLLMs) have achieved remarkable success across diverse real-world tasks~\cite{liu2023visual,yin2024survey,zeng2025avam, li2026diffgraph}. 
However, these models are usually trained on massive web-scraped data with limited filtering, often including toxic, offensive, and even dangerous information (e.g., technical details of constructing explosive devices), which makes these models prone to generating unsafe responses~\cite{ouyang2022training, gehman2020realtoxicityprompts, rafailov2023direct}. 
To avoid harmful responses, practitioners typically employ safety alignment techniques to align the behavior of large models with safe and harmless outputs ~\cite{ouyang2022training, rafailov2023direct}. 

However, even with alignment, LLMs and MLLMs can still be ``attacked'', i.e., their safeguards could still be bypassed by malicious intent, leading them to produce harmful outputs~\cite{liu2024mm,zou2023gcg}. 
This kind of attack is called the ``jailbreak'' attack, which poses non-trivial safety risks and has gained much research attention recently~\cite{ yang2025distraction, jeong2025playing, Hao_2025_CVPR}.
Generally, the jailbreak research focuses on intentionally simulating real-world jailbreak attacks, causing large models to produce harmful outputs, through which we can uncover potential vulnerabilities and societal risks in the models and thus further develop corresponding defense mechanisms.

\begin{figure}[t]
\centering
\includegraphics[width=1\linewidth]{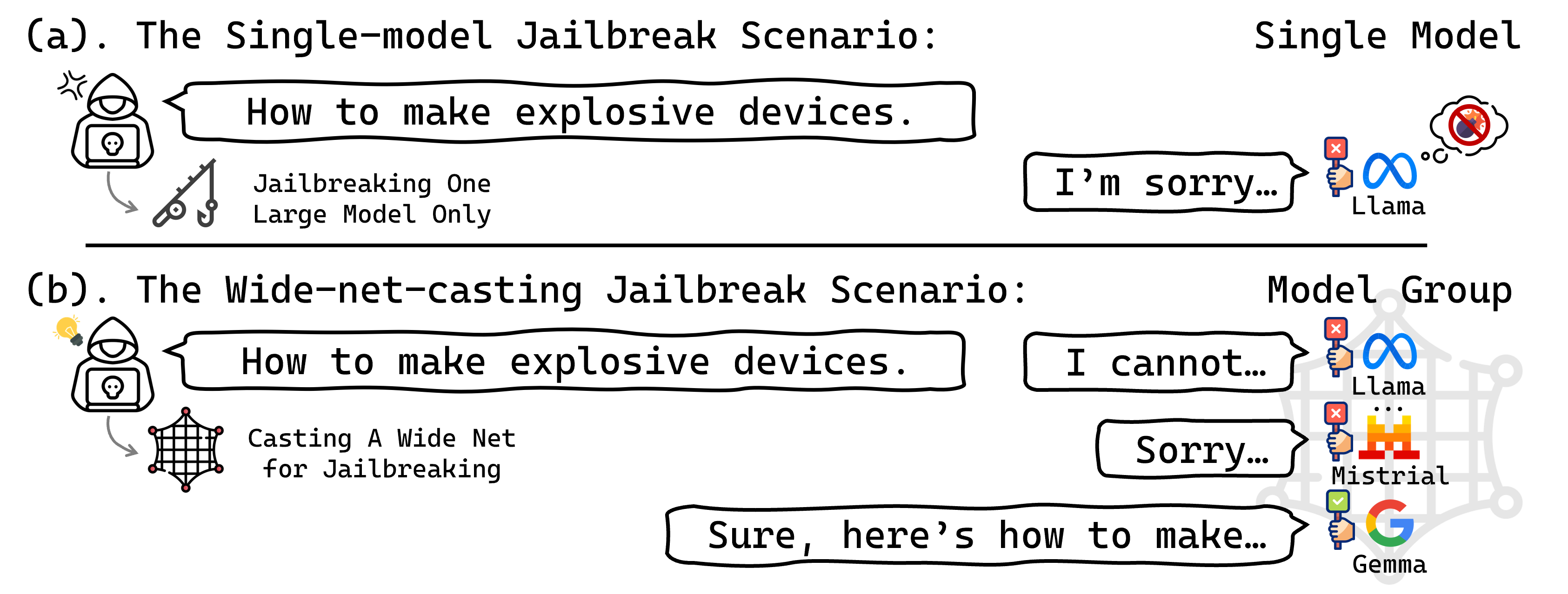}
\caption{Illustration of the single-model jailbreak scenario and the wide-net-casting jailbreak scenario. As shown, in the wide-net-casting scenario, unlike the single-model case, successfully jailbreaking any one large model in the group is sufficient for the adversary to obtain a desired harmful response.} 
\label{fig:1}
\end{figure}

 In this context, many jailbreak methods have been proposed 
\cite{qi2024visual, li2024images, paulus2024advprompter}.
However, prior research has largely overlooked a potential jailbreak attack scenario, \textit{the wide-net-casting scenario}. Such a potential scenario can be inspired by people's typical daily usage patterns of large models. Specifically, in the real world, when users send a complex query (e.g., a complex math problem) to a large model and receive an unsatisfactory response, we may tend to switch to another large model. Given availabilities of various publicly available large models 
(e.g.,  Llama~\cite{grattafiori2024llama}, Mistral~\cite{jiang2023clip}, and Gemma~\cite{team2024gemini}), users can ``cast a wide net'' over a group of multiple large models that exhibit various strengths and domain specializations, issuing the same request to various models to seek a correct and accurate answer.

This observation leads to an important hypothesis: malicious users could also employ similar strategies to achieve their desired harmful outputs under the wide-net-casting scenario (in contrast to the single-model scenario that focuses on jailbreaking one large model only). 
Given that large models within different families exhibit distinct characteristics, they usually possess distinct, model-specific vulnerabilities. 
Consequently, when multiple such large models are grouped and simultaneously attacked, their collective vulnerabilities can be amplified, thereby increasing the overall risk of successful jailbreak. 
For instance, as  shown in Fig.~\ref{fig:1}, if one large model (in the model group) fails to provide responses on how to build explosive devices, attackers can easily switch to other large models and thus succeed as long as any single model provides the technical details of building explosive devices. 
This indicates that risks of jailbreaking in the wide-net-casting jailbreak scenario may be higher, compared to those reported in prior studies under the single-model setting. 
Yet, despite its potential to expose significant risks, this practical jailbreak scenario remains unexplored.

Recognizing this critical gap, this work aims to systematically investigate the safety risks posed by the wide-net-casting jailbreak scenario, from two key aspects. 
\textbf{Aspect 1} examines whether directly adapting existing single-model jailbreak attacks to this scenario can significantly amplify safety risks when targeting a group of large models from different families. It also analyzes how these safety risks vary (i) when the target models belong to the same family (which thus may share similar vulnerabilities), and (ii) when additional safeguards beyond each model’s original safety alignment are applied.
\textbf{Aspect 2} simulates skilled adversaries who explicitly recognize this scenario and craft attacks tailored to it, evaluating the resulting safety risks.

For \textbf{Aspect 1}, our analysis (Sec.~\ref{sec: analysis}) reveals   directly adapting single-model jailbreaks to the wide-net-casting scenario can already lead to substantial safety risk amplification. 
For example, on the AdvBench dataset~\cite{zou2023gcg}, adapting the single-model GCG jailbreak method~\cite{zou2023gcg} to the wide-net-casting scenario increases the jailbreak success rate by at least 28.8\%, from 46.2\% to 75.0\% (see the first row of results in Tab.~\ref{tab:4}).
This amplification trend persists even when targeting large models from the same family or employing additional safeguards.
For \textbf{Aspect 2}, to simulate skilled adversaries, inspired by exploration-to-exploitation transition in optimization~\cite{perera2024annealed, xu2025annealed}, we design a novel jailbreak method tailored to the wide-net-casting scenario. The method pairs each target large model with a dedicated  ``jailbreak expert'', encouraging each expert to focus exclusively on its paired model’s distinct, model-specific vulnerabilities rather than attempting to cover all weaknesses uniformly. This targeted specialization prevents over-spreading each expert’s attention and ensures that each large model’s vulnerabilities are attacked by a dedicated expert, increasing the chance that at least one large model produces harmful outputs (detailed in Sec.~\ref{sec: new attack}). 
Empirically, our tailored attack can even achieve 100\% jailbreak success rates in some experiments when the target large models rely solely on their original safety alignments, exposing a stark and previously unexplored safety risk: \textbf{when adversaries optimize for the wide-net-casting scenario, large models can become alarmingly fragile}.

Our contributions are: 1) 
We are the first to reveal the previously unexplored \textit{ wide-net-casting jailbreak scenario}, and through comprehensive analysis, we uncover its previously overlooked safety risks. 
2) As a key part of analysis, we propose a novel jailbreak method tailored to this scenario, thereby more comprehensively exposing the underlying risks of wide-net-casting attacks.

\section{Related Work}
\label{sec: related}

Recent studies have shown that both LLMs and MLLMs are vulnerable to jailbreak attacks~\cite{Hao_2025_CVPR, geng2025instruction, shayegani2023jailbreak, liu2023autodan, xie2024jailbreaking}. Among various jailbreak strategies, optimization-based jailbreaks~\cite{liu2023autodan, yang2025distraction, jeong2025playing, Hao_2025_CVPR, geng2025instruction, paulus2024advprompter} have emerged as a mainstream approach, demonstrating remarkable empirical effectiveness~\cite{zhou-etal-2024-virtual, jia2025improved}. In this work, aiming to maximize the exposure of safety risks under the wide-net-casting scenario, we then primarily focus on optimization-based jailbreaks. Such methods can be broadly categorized into instance-based and model-based approaches: the former directly optimizes individual adversarial samples, while the latter learns an adversarial sample generator (e.g., a text-suffix generator for LLMs or an adversarial image generator for MLLMs). For \textbf{instance-based approaches}, targeting LLMs, GCG~\cite{zou2023gcg} optimizes text suffixes for harmful prompts, while AutoDan~\cite{liu2023autodan} employs genetic search to automate stealthy jailbreaks
Targeting MLLMs, Qi et al.~\cite{qi2024visual} optimizes adversarial images using a prompt-tuning-inspired approach, 
MLAI~\cite{Hao_2025_CVPR} iteratively applies gradient-based updates to improve a scenario-aware initial image for attacking,
and HADES~\cite{li2024images} embeds harmful intent into images via gradient refinement. 
In addition to instance-based methods, recently, a growing body of \textbf{model-based approaches}~\cite{liao2024amplegcg, sun2024iterative, paulus2024advprompter} have been further proposed. Typically, model-based approaches curate high-quality adversarial samples from instance-based attacks and use them to supervise the adversarial generator, achieving state-of-the-art jailbreak performance across diverse setups~\cite{sun2024iterative, xie2024jailbreaking}. Remiss~\cite{xie2024jailbreaking} searches for effective jailbreak suffixes, and uses these suffixes to train an adversarial sample generator.
Different from existing jailbreak methods that are typically designed for single-model jailbreak, this work introduces, for the first time, the wide-net-casting jailbreak scenario, a practical yet previously unexplored setting.

\section{Risk Analysis of Wide-net-casting Jailbreak}
\label{sec: analysis}
To investigate risks underlying the wide-net-casting scenario, we first note that existing jailbreak methods, both instance-based and model-based, are typically designed for single models. 
In existing instance-based methods, given a testing harmful intent and a target large model, the attacker typically first optimizes the intent into an adversarial sample specifically against that model, and then uses the optimized sample to jailbreak the same model. 
In existing model-based methods, given a training set of harmful intents and a target large model, the attacker uses all these intents to train an adversarial sample generator so that, at test time, the trained generator can automatically produce adversarial samples from arbitrary harmful intents to jailbreak the target model.
Since both types of methods inherently target single models, and no method has been developed specifically for wide-net-casting jailbreaks, a natural first step to understand the safety risks underlying the wide-net-casting scenario can be to examine how existing single-model methods behave when they are directly adapted to this new scenario.

\begin{figure}[t]
  \centering
   \includegraphics[width=1\linewidth]{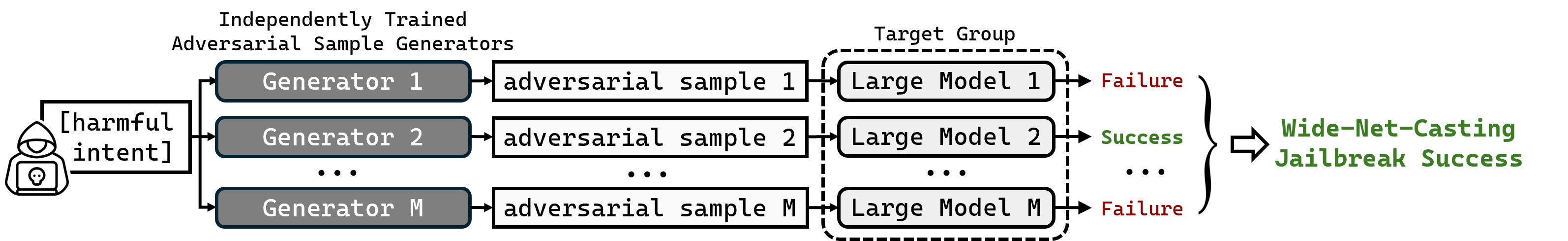}
   \caption{Illustration of straightforward adaptation of an existing model-based jailbreak method to the wide-net-casting scenario. A jailbreak attack is deemed successful in the wide-net-casting scenario as long as any target large model is successfully jailbroken.}
   \label{fig:2}
\end{figure}

To this end, we find that existing jailbreak methods, both instance-based and model-based, can be adapted to the wide-net-casting scenario in a straightforward manner.
To adapt an instance-based method, given a harmful intent and a group of $M$ target large models, for each large model, we apply an independent copy of the method to that model, optimizing the given intent into a distinct adversarial sample specifically crafted to jailbreak that model.
We then record whether any of these samples successfully breach its respective target. 
To adapt a model-based method, we instantiate $M$ independent copies of the adversarial sample generator and train each copy independently on the full training set of harmful intents, with each generator optimized against a different target large model.
During testing, given a harmful intent, as shown in Fig.~\ref{fig:2}, we run all $M$ trained generators in parallel, use each generator to produce a separate adversarial sample, and feed each sample to its corresponding large model to attempt a jailbreak attack. Similar to the instance-based case, we then record whether any of these attacks succeed to determine jailbreak success under the wide-net-casting scenario.
Through the above straightforward adaptation processes, existing single-model jailbreak techniques, both instance- and model-based, can be leveraged to expose the potential safety risks introduced by wide-net-casting. Building on this foundation, our investigation then centers on three key questions:

{\bf Q1:} When existing single-model jailbreak attacks are adapted to the wide-net-casting scenario, does this scenario amplify safety risks compared to the single-model scenario? If so, to what extent? {\bf Q2:} Given that large models within the same family often share similar architectures or training data and may therefore exhibit fewer model-specific vulnerabilities, how do associated risks change when wide-net-casting attacks are performed exclusively on models from the same family? {\bf Q3:} When large models are equipped with additional safeguards beyond their original safety alignment mechanisms, how does the overall risk level vary?

\subsection{Empirical Analysis for Q1} 
\label{sec: protocol_a}

\noindent\textbf{Datasets.} We use two widely used datasets, AdvBench~\cite{zou2023gcg} and MM-SafetyBench~\cite{liu2024mm}. In our analysis, we test LLMs on AdvBench and test MLLMs on MM-SafetyBench. We follow~\cite{paulus2024advprompter, xie2024jailbreaking} for the train-test split of AdvBench, and follow~\cite{Hao_2025_CVPR} for the train-test split of MM-SafetyBench.

\noindent\textbf{Evaluation Metrics.} To assess attack success, we report the Attack Success Rate (ASR) metric~\cite{zou2023gcg} when jailbreaking a single large model. However, this metric evaluates each model independently and thus fails to capture the wide-net-casting scenario, where an attack is considered successful if any model within a group is jailbroken.  To address this, we further introduce the Wide-net-casting Attack Success Rate (WASR) metric: $\text{WASR}=\frac{1}{N}\sum_{n=1}^{N}\bigvee_{m=1}^{M}s_m^n$, where $N$ is the number of harmful intents used in evaluation, and $M$ is the number of large models in the group. The binary variable $s_m^n \in \{0,1\}$ indicates whether the $n$-th intent successfully jailbreaks the $m$-th model in the group (1 for success, 0 for failure). The logical OR operator $\bigvee$ aggregates these binary variables across all $M$ models, yielding $\bigvee_{m=1}^{M}s_m^n = 1$ if at least one model in the group is successfully jailbroken by the $n$-th intent, and yielding 0 otherwise. In this way, WASR quantifies the proportion of harmful intents that successfully jailbreak at least one model within the target group, effectively measuring jailbreak success under the wide-net-casting scenario. 
Following~\cite{Hao_2025_CVPR}, we use Beaver-Dam-7B~\cite{ji2023beavertails} as the judge model to determine whether each individual large model has been successfully jailbroken or not, based on the jailbreak quality rating that Beaver-Dam-7B assigns to the large model’s response. 
We report WASR as the primary evaluation metric under the wide-net-casting scenario.

Besides, following~\cite{wang2025steering, ziqi-etal-2025-visual}, we also evaluate the response toxicity in jailbreaking and adopt the Toxicity Score metric~\cite{wang2025steering}, which can indicate severity of jailbreak outputs~\cite{bachu2025layerwise}. Yet, while suitable for single-model jailbreaks, this metric evaluates each model’s response independently and is thus not directly applicable to wide-net-casting jailbreaks. To address this, we introduce the Wide-net-casting Toxicity Score (W-Toxicity Score). 
Specifically, for each harmful intent, among the responses generated by models in the target group, we simulate attackers to select the response most likely to be used for harm using a large model, and define the W-Toxicity Score as the toxicity score of this selected response. We find that using different large models can yield highly consistent selections (details in the Appendix~\ref{extend_sec3}).

\begin{table}[t]
  \caption{Evaluation of jailbreaking LLMs across different model families under the wide-net-casting scenario on AdvBench.}
  \centering
  \label{tab:4}
        \resizebox{\columnwidth}{!}{
        \fontsize{40}{50}\selectfont
        \setlength{\heavyrulewidth}{0.10em}
        \setlength{\lightrulewidth}{0.05em}
        \setlength{\cmidrulewidth}{0.05em}
        \begin{tabular}{@{}lcccccc@{}}
          \toprule
          \multirow{2}{*}{Defense} & \multirow{2}{*}{Attack\space\space\space} &
          \multicolumn{4}{c}{ASR / Toxicity Score} &
          \multirow{2}{*}{\makecell{WASR / W-Toxicity\\ Score}} \\
          \cmidrule(lr){3-6}
          & & Gemma-2-9b & Vicuna-7b-v1.5 & Llama-3.1-8b & Mistral-7b & \\
          \midrule
          \multirow{2}{*}{\makecell[l]{Original \\ Safety Alignment}}
            & GCG    & 24.0\% / 0.221  & 36.5\% / 0.328 & 28.8\% / 0.241  & 46.2\% / 0.443 & \textbf{75.0\% / 0.736} \\
            & ReMiss & 45.2\% / 0.370  & 49.0\% / 0.424 & 38.5\% / 0.352  & 86.5\% / 0.811 & \textbf{92.3\% / 0.877} \\
          \midrule
          \multirow{2}{*}{\makecell[l]{+ SmoothLLM\\~\cite{robey2023smoothllm}}}
            & GCG    & 11.5\% / 0.081  & 17.3\% / 0.124 & 14.4\% / 0.098  & 26.9\% / 0.202 & \textbf{46.1\% / 0.353} \\
            & ReMiss & 21.2\% / 0.187  & 27.9\% / 0.260 & 18.3\% / 0.155  & 38.5\% / 0.389 & \textbf{61.5\% / 0.530} \\
          \midrule
          \multirow{2}{*}{\makecell[l]{+ RobustKV\\~\cite{jiang2024robustkv}}}
            & GCG    & 5.9\% / 0.021   & 16.4\% / 0.121 & 7.9\% / 0.038   & 24.6\% / 0.228 & \textbf{37.3\% / 0.325} \\
            & ReMiss & 10.8\% / 0.059  & 28.7\% / 0.224 & 15.3\% / 0.107  & 34.4\% / 0.289 & \textbf{56.1\% / 0.511} \\
          \bottomrule
        \end{tabular}
        }
\end{table}

\begin{table}[t]
\caption{Evaluation of jailbreaking MLLMs across different model families under the wide-net-casting scenario on MM-SafetyBench.}
\centering
\small
\resizebox{\columnwidth}{!}{
\fontsize{40}{50}\selectfont
\setlength{\heavyrulewidth}{0.10em}
\setlength{\lightrulewidth}{0.05em}
\setlength{\cmidrulewidth}{0.05em}
\begin{tabular}{@{}lcccccc@{}}
\toprule
\multirow{2}{*}{Defense} & \multirow{2}{*}{Attack} &
\multicolumn{4}{c}{ASR / Toxicity Score} & \multirow{2}{*}{\makecell{WASR / W-Toxicity \\Score}} \\
\cmidrule(lr){3-6}
 & & MiniGPT-4-13b & LLaVA-1.5-13b & InstructBLIP & Qwen2-VL-7B&  \\
\midrule
\multirow{2}{*}{\makecell[l]{Original \\Safety Alignment}}  
  & MLAI         
    & 77.7\% / 0.732 & 64.1\% / 0.602 & 50.5\% / 0.455 & 38.7\% / 0.349 & \textbf{89.1\% / 0.824} \\
  & MLAI+ PixArt-\textalpha
    & 77.9\% / 0.744 & 65.2\% / 0.609 & 53.3\% / 0.472 & 39.7\% / 0.355 & \textbf{93.7\% / 0.891} \\
\midrule
\multirow{2}{*}{\makecell[l]{+ VLGuard\\~\cite{zong2024safety}}}  
  & MLAI         
    & 22.6\% / 0.182 & 13.1\% / 0.092 & 19.3\% / 0.152 & 11.6\% / 0.091 & \textbf{36.2\% / 0.310} \\
  & MLAI+ PixArt-\textalpha
    & 22.9\% / 0.198 & 13.7\% / 0.108 & 20.1\% / 0.177 & 12.5\% / 0.096 & \textbf{40.2\% / 0.387} \\
\midrule
\multirow{2}{*}{\makecell[l]{+ IMMUNE\\~\cite{ghosal2025immune}}} 
  & MLAI         
    & 18.4\% / 0.139 & 4.7\% / 0.025 & 17.6\% / 0.126 & 8.7\% / 0.039 & \textbf{33.6\% / 0.301} \\
  & MLAI+ PixArt-\textalpha
    & 20.9\% / 0.151 & 6.2\% / 0.038 & 18.1\% / 0.137 & 9.1\% / 0.050 & \textbf{37.2\% / 0.321} \\
\midrule
\multirow{2}{*}{\makecell[l]{+ ASTRA\\~\cite{wang2025steering}}} 
  & MLAI         
    & 10.1\% / 0.061 & 11.2\% / 0.078 & 13.7\% / 0.083 & 6.1\% / 0.038 & \textbf{29.5\% / 0.257} \\
  & MLAI+ PixArt-\textalpha
    & 10.9\% / 0.070 & 12.3\% / 0.091 & 15.9\% / 0.099 & 7.1\% / 0.044 & \textbf{32.9\% / 0.271} \\
\bottomrule
\end{tabular}
}
\label{tab:5.1}
\end{table}

\noindent\textbf{Jailbreak Attacks and Target Models.} We begin by introducing the target models selected for jailbreaking LLMs. For comprehensive evaluation under the wide-net-casting scenario, we consider four widely-used LLMs as our targets: Gemma-2-9b~\cite{team2024gemini}, Vicuna-7b-v1.5~\cite{zheng2023judging}, Llama-3.1-8b~\cite{grattafiori2024llama}, and Mistral-7b~\cite{jiang2023clip}. Each of these LLMs is widely used as a target model in prior jailbreak research for evaluation~\cite{zou2023gcg, liu2023autodan, lin2024towards}. For attack methods, we adopt two representative approaches: a classic instance-based LLM jailbreak method GCG~\cite{zou2023gcg}, and a state-of-the-art model-based LLM jailbreak method ReMiss~\cite{xie2024jailbreaking}.

For MLLMs, we consider four widely used MLLMs as attack targets: MiniGPT-4-13b~\cite{zhu2023minigpt}, LLaVA-1.5-13b~\cite{liu2023visual}, InstructBLIP~\cite{dai2023instructblip}, and Qwen2-VL-7B~\cite{wang2024qwen2}. For attack methods, we first adopt an instance-based jailbreak method MLAI~\cite{Hao_2025_CVPR}. We further aim to also explore model-based jailbreaks for MLLMs under the wide-net-casting scenario. However, model-based jailbreaks for MLLMs remain largely underexplored. To fully expose risks in the wide-net-casting scenario, we then also construct a model-based jailbreak for MLLMs, via adapting LLM-oriented model-based pipelines~\cite{liao2024amplegcg, sun2024iterative, paulus2024advprompter, xie2024jailbreaking}. Specifically, we first leverage MLAI~\cite{Hao_2025_CVPR} to generate high-quality adversarial images. Inspired by~\cite{li2024images, wu2025realistic}, we then adopt PixArt-$\alpha$~\cite{chen2024pixart} as the adversarial image generator and fine-tune it on these samples. The resulting method, denoted as ``MLAI~\cite{Hao_2025_CVPR} + PixArt-$\alpha$~\cite{chen2024pixart}'', serves as our model-based jailbreak for MLLMs. Additional details w.r.t. this method are in Appendix~\ref{m+p}.

\noindent\textbf{Experimental Results.} As shown in Tab.~\ref{tab:4} and Tab.~\ref{tab:5.1} (``Original Safety Alignment'' parts), across both LLMs and MLLMs and on different datasets, adapting existing single-model jailbreaks to the wide-net-casting scenario consistently increases both attack success rates and response toxicity. These results affirmatively answer \textbf{Q1}: \ul{Adapting existing single-model jailbreak attacks to the wide-net-casting scenario can consistently and significantly amplify safety risks.}
This amplification occurs as different large models often exhibit model-distinct vulnerabilities: a jailbreak attack that fails on one model may succeed on another. In the wide-net-casting scenario, these model-specific vulnerabilities can be collectively exploited, thereby amplifying safety risks. We also conducted experiments of jailbreaking MLLMs on AdvBench, see Appendix~\ref{extend_sec3} for details.

\begin{table}[t]
\caption{Evaluation of jailbreaking MLLMs within the same model family under the wide-net-casting scenario on MM-SafetyBench.}
\centering
\small
\resizebox{\columnwidth}{!}{
\fontsize{40}{50}\selectfont
\setlength{\heavyrulewidth}{0.10em}
\setlength{\lightrulewidth}{0.05em}
\setlength{\cmidrulewidth}{0.05em}
\begin{tabular}{@{}lccccc@{}} 
\toprule 
 \multirow{2}{*}{Method} & 
\multicolumn{4}{c}{ASR / Toxicity Score} & \multirow{2}{*}{\makecell{WASR / W-Toxicity \\Score}} \\ 
\cmidrule(lr){2-5} 
  & LLaVA-1.5-13b\space\space & LLaVA-1.6-vicuna-13b\space\space & LLaVA-1.6-vicuna-7b\space\space & LLaVA-llama2-13b& \\ 
\midrule 
 MLAI &
 64.1\% / 0.602 & 16.7\% / 0.135 & 34.3\% / 0.291 & 75.2\% / 0.708 & \textbf{87.6\% / 0.832 } \\ 
 MLAI+PixArt-\textalpha &
 65.2\% / 0.609 & 18.2\% / 0.151 & 37.6\% / 0.327 & 79.3\% / 0.746 & \textbf{89.9\% / 0.865 } \\ 
\bottomrule
\end{tabular}
}
\label{stab: y}
\end{table}

\begin{table}[t]
\caption{Evaluation of jailbreaking MLLMs within the same model family and same version on MM-SafetyBench.}
\centering
\small
\resizebox{0.8\columnwidth}{!}{
\fontsize{40}{50}\selectfont
\setlength{\heavyrulewidth}{0.10em}
\setlength{\lightrulewidth}{0.05em}
\setlength{\cmidrulewidth}{0.05em}
\begin{tabular}{@{}lccc@{}}
\toprule
 \multirow{2}{*}{Method} &
\multicolumn{2}{c}{ASR / Toxicity Score} & \multirow{2}{*}{WASR / W-Toxicity Score}\\
\cmidrule(lr){2-3}
  & LLaVA-1.6-vicuna-13b\space\space\space &  LLaVA-1.6-vicuna-7b & \\
\midrule
   MLAI  &16.7\% / 0.135  & 34.3\% / 0.291 & \textbf{38.9\% / 0.352 } \\
   MLAI+PixArt-\textalpha  &18.2\% / 0.151 & 37.6\% / 0.327 & \textbf{41.1\% / 0.372 } \\
\bottomrule
\end{tabular}
}
\label{stab: x}
\end{table}

\subsection{Empirical Analysis for Q2}
\label{sec: Q2}

\noindent\textbf{Experimental Setups.} We keep the evaluation metrics and jailbreak methods consistent with Sec.~\ref{sec: protocol_a}, varying only the target models by restricting to models from the same family. 
For MLLMs, we build target groups from the popular LLaVA family on MLLMs, initially including four representative models (see Tab.~\ref{stab: y}). For LLMs, we construct experiments on the Llama family in Appendix~\ref{extend_sec3} for details. To further analyze risks among even more closely related models, we construct another group comprising LLaVA models from both the same family and version (see Tab.~\ref{stab: x}).

\noindent\textbf{Experimental Results.} As shown in Tab.~\ref{stab: y} and Tab.~\ref{stab: x}, even when the target group is limited to models within the same family, their (possibly fewer) model-specific vulnerabilities can still be exploited in the wide-net-casting scenario, and the resulting amplification of safety risks remains evident.
Moreover, Tab.~\ref{stab: x} shows that while amplification becomes
less pronounced when the target group is further restricted to models of the same family and version, a modest increase in vulnerability still persists. These results answer \textbf{Q2}: \emph{The wide-net-casting scenario consistently amplifies safety risks for large models, even when the attack is performed exclusively on models from the same family.}

\subsection{Empirical Analysis for Q3}
\label{sec: q3}

\noindent\textbf{Experimental Setups.} We largely follow the setups in Sec.~\ref{sec: protocol_a}, differing only by equipping the large models with additional safeguards. We  evaluate diverse additional safeguards.
For LLMs, we consider two representative approaches: the classic SmoothLLM~\cite{robey2023smoothllm} and recent RobustKV~\cite{jiang2024robustkv}. For MLLMs, we test three representative safeguards: the classic VLGuard~\cite{zong2024safety}, and two recent ones, IMMUNE~\cite{ghosal2025immune} and ASTRA~\cite{wang2025steering}.

\noindent\textbf{Experimental Results.} As shown in Tab.~\ref{tab:4} and Tab.~\ref{tab:5.1}, though equipping additional safeguards substantially reduces the (single-model) ASR of large models, the WASR remains noticeably higher than the ASR across LLMs and MLLMs, suggesting that the wide-net-casting scenario can still exploit model-specific vulnerabilities even under additional safeguards. These results answer \textbf{Q3}: \emph{Even with additional safeguards, the wide-net-casting scenario can still consistently and substantially elevate safety risks.}

\section{Risk Exposure Under Skilled Adversaries}
\label{sec: new attack}

In Sec.~\ref{sec: analysis}, we examine the wide-net-casting scenario under an independently-optimized threat manner, independently optimizing adversarial samples (or their generators) to jailbreak each large model. The analysis already shows that this previously overlooked scenario can consistently elevate safety risks across diverse experimental setups, revealing a distinct, high-risk paradigm that merits dedicated attention in future evaluation and defense research. However, because multiple large models co-exist within a target group in the wide-net-casting scenario, independently-optimized attacks may still underestimate the true threat level of this scenario.
A skilled adversary could instead exploit cross-model information to craft coordinated jailbreaks that jointly target multiple models, thereby further amplifying attack effectiveness. Motivated by this insight and further drawn inspiration from exploration-to-exploitation transition in optimization~\cite{perera2024annealed},
in this section, we simulate such skilled adversaries by developing a new jailbreak method explicitly tailored to the wide-net-casting scenario, aiming to more comprehensively expose the underlying risks.

\subsection{A Jailbreak Tailored to Wide-net-casting}
\label{A New}

We observe in Sec.~\ref{sec: analysis} that when single-model jailbreak methods are straightforwardly adapted to the wide-net-casting scenario, model-based approaches consistently outperform instance-based ones across diverse experimental setups. 
Based on this, we adopt the model-based paradigm as the foundation for developing a new jailbreak method tailored to the wide-net-casting scenario,  aiming to more thoroughly reveal the underlying safety risks of this scenario.
Pursuing this direction, we first notice that the straightforward adaptation of existing model-based approaches to wide-net-casting can suffer from a key limitation, namely, a \textit{lack of specialization} within each adversarial sample generator. 

Specifically, as detailed in Sec.~\ref{sec: analysis}, under such an adaptation, given a group of $M$ target large models, the $M$ corresponding adversarial sample generators are each trained independently on the entire training set of harmful intents, pushing every generator to attempt to cover all weaknesses. While such an all-covering strategy can be appropriate in single-model settings where only one generator targeting a single large model typically exists, it can be ill-suited to wide-net-casting, where $M$ generators co-exist and an attack succeeds once any generator successfully breaches its corresponding large model (as shown in Fig.~\ref{fig:2}). Thus, it is unnecessary for every generator to attempt to fully cover all weaknesses. Instead, given that each target large model often exhibits distinct, model-specific vulnerabilities, a more effective jailbreak strategy in the wide-net-casting scenario shall be to optimize all $M$ generators jointly with an emphasis on \textit{specialization}, focusing each generator exclusively on vulnerabilities of its corresponding large model rather than on all weaknesses. This emphasis on specialization prevents over-dispersion of each generator’s attention and ensures that every large model is attacked by a specialized (dedicated) expert, increasing likelihood that at least one large model will be breached under wide-net-casting jailbreaks.

Building on the advantages of specialization discussed above, we explore how to effectively equip each adversarial sample generator with such specialization. Our key intuition is that specialization can be encouraged by driving each generator to learn from intents that it already handles relatively well: by repeatedly reinforcing its existing strengths, a generator’s expertise can be progressively refined, thus sharpening its specialization. 
The remaining question, then, is how to determine, for each training intent, which generator among the $M$ generators already handles that intent relatively well (i.e., already relatively proficient at it).
Given this, we observe that, given a training intent, the loss value calculated from a generator's jailbreak attempt over its corresponding large model  can reflect its proficiency on that intent: a smaller loss value  can indicate higher proficiency.

Based on this insight while also drawn inspiration from \cite{guzman2012multiple}, some naive strategies using such loss values as proficiency indicators first come to our mind, such as the two described below:
(1) \textit{Naive Strategy 1}: Recall that after the analysis in Sec.~\ref{sec: analysis}, we have obtained a set of independently-trained adversarial sample generators that, though trained in isolation, each capture some meaningful jailbreak-related knowledge. Given $M$ such generators corresponding to a target group of $M$ large models, we can then use their loss values to quantify proficiency: each training intent is passed through all generators, and the one yielding the smallest loss is deemed the most proficient.  Next, to reinforce each generator’s strengths and thereby enhance its specialization, we can assign each training intent to its most proficient generator, and subsequently fine-tune each generator exclusively on its assigned intents. (2) \textit{Naive Strategy 2}: Instead of relying on loss values from independently-trained generators, we can also use the loss values computed at each iteration of the subsequent (joint) training process as proficiency indicators. Specifically, at each training step (iteration), we can pass the current intent through all generators and update only the one with the smallest loss. This update scheme shall also continually reinforce each generator’s expertise and progressively enhance specialization.

Yet, despite appearing promising, both above naive strategies share a common limitation: they quantify proficiency from \textit{intermediate} loss values, either those derived from independently-trained generators or those measured during joint training iterations. Yet, these intermediate losses may not align with losses produced by an ``ideally'' well-trained generator that has achieved ``ideal'' specialization, since the optimization landscape of each adversarial sample generator, being deep and highly non-convex, is inherently very complex. 
Consequently, generators showing relatively large intermediate losses may still achieve very small loss after ``ideal'' specialization. In other words, intermediate losses serve only as \textit{noisy indicators} of the true proficiency that would emerge after ``ideal'' specialization.
Relying entirely on them for update scheduling may therefore misguide the training process (e.g., by overlooking the generator which should actually be updated for the current intent), resulting in each generator becoming only sub-optimally specialized. As a result, these naive strategies may still underestimate actual risks inherent in the wide-net-casting scenario. 

To handle the above limitation, since the loss value of an ``ideally'' specialized generator is clearly inaccessible during training, we draw inspiration from optimization theory on how classical optimization is conducted when only a \textit{noisy, instantaneous indicator} of the true objective is available. To this end, we find that prior studies~\cite{van1987simulated, shi1999empirical, perera2024annealed, shi1998modified, bengio2015scheduled} suggest a key remedy: structuring the optimization as a gradual transition from \textit{exploration} to \textit{exploitation}. Specifically, these studies suggest that, when the true objective cannot be directly observed, to well-optimize towards the objective, \ul{each optimization step should aim to maximize exploitation while maintaining at least a non-zero and monotonically decreasing level of sustainable exploration}. In our problem, \textit{exploitation} refers to treating the intermediate loss as a fully reliable indicator and concentrating updates as much as possible on generators with smaller intermediate losses, whereas \textit{exploration} refers to allowing each generator to retain some 
opportunity for updates even when it performs relatively poorly on the current training intent. Building on this interpretation, we note that, by jointly optimizing the $M$ adversarial sample generators in the above-underlined manner, we shall be able to simultaneously encourage effective specialization within each generator (by maximizing exploitation), while explicitly accounting for the noise inherent in intermediate loss values (by maintaining sustainable exploration).
This motivates us to design our training process to operate precisely in this exploration–exploitation manner.

After conceptually defining our goal as above (see the underlined sentence), we next formulate it mathematically to enable its practical achievement. This conceptual-to-practical transition, however, is non-trivial.  To accomplish it, we first decompose the overall goal into sub-goal \circledone: \textit{maximizing exploitation}, and sub-goal \circledtwo: \textit{maintaining sustainable exploration}, and express each in analytical form. We then integrate these sub-goals into the complete formulation.

\ul{For the sub-goal \circledone~(\textit{maximizing exploitation})}, recall that its objective is to concentrate updates as much as possible on generators with smaller losses. Let $\bm{\ell_t}=(\ell^1_t,\ldots,\ell^M_t)$ denote the training losses of the $M$ generators at a certain training step $t$. Let $\bm{\eta_t}=(\eta^1_t,\ldots,\eta^M_t)$ be a weight vector at a certain training step $t$ with $\eta^m_t \ge 0$ and $\sum_{m=1}^M \eta^m_t = 1$, where $\eta^m_t$ represents the relative update weight assigned to the $m$-th generator in the update schedule. Using these notations, the sub-goal \circledone~can be viewed as a problem that aims to simultaneously satisfy two criteria: (1) generators with smaller losses $\ell^m_t$ should be assigned larger update weights $\eta^m_t$ as much as possible; and (2) \bm{$\eta_t$} needs to remain a valid weight vector in the simplex $\Delta_M$, i.e., $\eta^m_t \ge 0$ and $\sum_{m=1}^M \eta^m_t = 1$. As theoretically shown in Appendix~\ref{Theoretical Analysis 1}, handling the problem with these criteria is equivalent to solving the following optimization problem:
\begin{equation}
\label{eq:global-loss}
\begin{aligned}
& \bm{\eta_t^*} \leftarrow \arg\min_{\bm{\eta_t}\in \Delta_M} \sum_{m=1}^M \eta_t^m \,\ell_t^m, \\ & ~\Delta_M=\{\bm{\eta_t}:\ \eta_t^m\ge0,\ \sum_{m=1}^M \eta_t^m=1\}.
\end{aligned}
\end{equation}
Therefore, the sub-goal \circledone~can be formally represented as the optimization problem in Eq.~\ref{eq:global-loss}.
\ul{Next, we formalize the sub-goal \circledtwo~(\textit{maintaining at least a non-zero and monotonically decreasing level of sustainable exploration})}, through the following three steps of reasoning. 
(1) Recall that in our setting, exploration refers to granting each generator some non-trivial opportunity for updates, even when it performs poorly on the current training intent. Accordingly, we find that, the level of exploration can actually be reflected by how \textit{spread-out} the update weights in $\bm{\eta_t}$ are: when $\bm{\eta_t}$ is highly concentrated (i.e., non-trivial weights are assigned to only few generators), exploration is low; conversely, when $\bm{\eta_t}$ is more uniformly distributed (i.e., non-trivial weights are broadly assigned), exploration is high. 
(2) We can then naturally quantify how \textit{spread-out} is $\bm{\eta_t}$ using its entropy~\cite{shannon1948mathematical}, defined as $H(\bm{\eta_t}) = -\sum_{m=1}^M \eta_t^m \log \eta_t^m$, where $0 \le H(\bm{\eta_t}) \le \log M$. A larger $H(\bm{\eta_t})$ indicates a more \textit{spread-out} (uniform) $\bm{\eta_t}$, and hence a higher level of exploration.
(3) Based on $H(\bm{\eta_t})$, we can formalize the sub-goal \circledtwo~as constraining $\bm{\eta_t}$ to satisfy, at each training step $t$,
\begin{equation}
\label{eq:exploration_constraint}
\begin{aligned}
& H(\bm{\eta_t}) \ge \bar{H}_t,
\end{aligned}
\end{equation}
where $\bar{H}_t = \log M \times \frac{T - t}{T}$ specifies a non-zero and monotonically decreasing entropy threshold, and $t \in \{0, \dots, T-1\}$ with $T$ denoting the total number of training steps.

Integrating Eq.~\ref{eq:global-loss} and Eq.~\ref{eq:exploration_constraint}, which respectively correspond to the two sub-goals, we can formulate our overall goal analytically as the following  constrained optimization problem:
\begin{equation}
\label{eq:entropy-constrained}
\begin{split}
\bm{\eta_t^*}  \leftarrow  &\arg\min_{\bm{\eta_t}\in\Delta_M}  \sum_{m=1}^M \eta_t^m \,\ell_t^m
\quad\text{s.t.}\quad H(\bm{\eta_t})\ \ge\ \bar{H}_t,\\
&\Delta_M=\{\,\bm{\eta_t}:\eta_t^m\ge 0,\ \sum_{m=1}^M \eta_t^m=1\,\}\,.
\end{split}
\end{equation}
Accordingly, achieving our goal now reduces to solving Eq.~\ref{eq:entropy-constrained}. Building on~\cite{perera2024annealed}, which solves constrained optimization problems using a Lagrangian formulation, Eq.~\ref{eq:entropy-constrained}, though seemingly complex, can be solved through the following four steps.

\textbf{Step (1).} We first rewrite \(H(\bm{\eta_t}) \ge \bar{H}_t\) into: $g(\bm{\eta_t})\triangleq \bar{H}_t - H(\bm{\eta_t}) = \bar{H}_t + \sum_{m=1}^M \eta_t^m \log \eta_t^m \le 0$. \textbf{Step (2).} We then introduce Lagrange multipliers~\cite{boyd2004convex}:
\(\beta_t \ge 0\) for the inequality \(g(\bm{\eta_t}) \le 0\);
\(\nu_t \in \mathbb{R}\) for the constraint \(\sum_{m=1}^M \eta_t^m = 1\); 
and \(\alpha^m_t \ge 0\) for the constraint \(\eta_t^m \ge 0\).
Denoting $\bm{\alpha_t} = (\alpha^1_t,\ldots,\alpha^M_t)$, the resulting Lagrangian is:
\begin{equation}
\label{eq: lagrangian}
\begin{split}
\mathcal{L}(\bm{\eta_t},\beta_t,\nu_t,\bm{\alpha_t})
=&\sum_{m=1}^M \eta_t^m \ell_t^m +\beta_t (\bar{H}_t+\sum_{m=1}^M \eta_t^m\log \eta_t^m) \\
& +\nu_t (\sum_{m=1}^M \eta_t^m-1)
-\sum_{m=1}^M \alpha_t^m \eta_t^m.
\end{split}
\end{equation}
\textbf{Step (3).} Applying the Karush-Kuhn-Tucker (KKT) first-order stationarity condition with respect to \(\eta_t^i\) then gives:
\begin{equation}
\label{eq:stationarity}
\frac{\partial\mathcal{L}}{\partial \eta_t^i} = \ell_t^i+\nu_t-\alpha^i_t+\beta_t (1+\log \eta_t^i) = 0.
\end{equation}
\textbf{ Step (4)}. Applying the complementary slackness condition (see Appendix~\ref{Theoretical Derivation 2} for details on its application), we can finally derive the optimal solution $\bm{\eta_t^*}$ to Eq.~\ref{eq:entropy-constrained} as:
\begin{equation}
\label{eq:final-eta}
\bm{\eta_t^*} = (\eta_t^{1, *}, \dots, \eta_t^{M, *}),~\eta_t^{i, *} = 
\frac{\exp(-\ell_t^i/\beta_t)}
{\sum_{m=1}^M \exp(-\ell_t^m/\beta_t)},
\end{equation}

which forms a Boltzmann distribution. Notably, at each training step $t$, based on $\bar{H}_t$ and  
losses $\bm{\ell_t} = (\ell_t^1,\ldots,\ell_t^M)$, $\bm{\eta_t^*}$ in Eq.~\ref{eq:final-eta} is uniquely determined,  since $\beta_t$ is uniquely specified by $\bar{H}_t$ and $\bm{\ell_t}$ (see Appendix~\ref{Theoretical 3} for proof). 

Overall, during training, at each training step $t$, based on $\bar{H}_t$ and $\bm{\ell_t}$ at this step, we first compute $\beta_t$, and then use $\beta_t$ together with $\bm{\ell_t}$ to derive $\bm{\eta_t^*}$ through Eq.~\ref{eq:final-eta}. With $\bm{\eta_t^*}$ determined, we subsequently update each of the $M$ adversarial sample generators according to $\bm{\eta_t^*}$, using the weighted loss $\eta_t^{m, *} \times \ell^m_t$ to update the $m$-th generator. As discussed earlier, arranging the update schedule in this manner according to $\bm{\eta_t^*}$ enables the generators to be optimized in a way that simultaneously maximizes exploitation while still maintaining sustainable exploration. Consequently, throughout training, each generator shall gradually evolve towards becoming a specialized expert in jailbreaking its corresponding large model (as elaborated in the front part of this subsection). This, in turn, increases the likelihood that at least one large model will be breached under wide-net-casting, thereby ultimately exposing the safety risks underlying the wide-net-casting jailbreak scenario in a more comprehensive manner.

\subsection{Overall Training and Testing}
\noindent\textbf{Training.} Given $M$ target large models, we first pair each model with an independently-trained adversarial sample generator. We then further jointly train these $M$ generators for $T$ steps, using the method in Sec.~\ref{A New}.
\noindent\textbf{Testing.} At test time, given a harmful intent, each generator converts the intent into an adversarial sample targeting its corresponding large model. Feeding these $M$ samples into their respective large models then yields $M$ candidate jailbreak outputs. Finally, we automatically select the highest-jailbreak-quality output among them as the final output of our method. Notably, this automatic selection process can be performed simply and accurately using a large model (e.g., Beaver-Dam-7B) or a template-based scoring method~\cite{lisafety}. See more details in Appendix~\ref{selection}.

\section{Evaluation of the Proposed Method}
\label{sec: exp}

\noindent\textbf{Experimental Setups \& Implementation Details.} To evaluate the effectiveness of our proposed jailbreak method tailored to the wide-net-casting scenario, we conduct two sets of experiments following the setups described in Sec.~\ref{sec: protocol_a} and Sec.~\ref{sec: q3}, respectively. 
For experiments targeting LLMs and MLLMs, we initialize the joint training process using adversarial sample generators independently trained by ReMiss~\cite{xie2024jailbreaking} and ``MLAI~\cite{Hao_2025_CVPR} + PixArt-$\alpha$~\cite{chen2024pixart}'', respectively (as in Sec.~\ref{sec: analysis}).
Then for both the LLM and MLLM experiments, the total number of joint training steps $T$ is set to 3,000. All experiments are conducted on four NVIDIA A100 GPUs. More implementation details are in Appendix~\ref{detail_sec5}.

\subsection{Main Results}
\label{sec: main results}

We compare our method with both the straightforward OR-aggregation adaptation of existing single-model jailbreak approaches, which serves as the baseline and is described in Sec.~\ref{sec: analysis}, and the two naive strategies tailored to the wide-net-casting scenario, introduced in Sec.~\ref{A New}.
For a fair comparison, both naive strategies use the same independently trained adversarial generators as ours and undergo an equal number of joint training steps.
Results targeting LLMs and MLLMs are reported in Tab.~\ref{tab:6} and Tab.~\ref{tab:7}, respectively. 

As shown, while the two naive strategies slightly outperform the straightforward adaptation of single-model methods, our approach consistently and very significantly outperforms the two naive strategies and the straightforward adaptation across datasets. 
Notably, across both LLMs and MLLMs, our method consistently achieves a 100\% jailbreak success rate when the target large models rely solely on their original safety alignments. The above shows the superior efficacy of our method.
 
\begin{table}[t]
\caption{Evaluation of jailbreaking LLMs using different methods tailored to the wide-net-casting jailbreak scenario.}
\centering
\resizebox{1\columnwidth}{!}{
\fontsize{40}{50}\selectfont
\setlength{\heavyrulewidth}{0.10em}
\setlength{\lightrulewidth}{0.05em}
\setlength{\cmidrulewidth}{0.05em}
\begin{tabular}{lcccc}
\toprule
\multirow{3}{*}{Dataset} &\multirow{3}{*}{Attack} & \multicolumn{3}{c}{WASR  / W-Toxicity Score}  \\
\cmidrule(lr){3-5}
& & \makecell{Original Safety Alignment} & \makecell{Original Safety Alignment\\  + SmoothLLM~\cite{robey2023smoothllm}} & \makecell{Original Safety Alignment\\  + RobustKV~\cite{jiang2024robustkv}}\\
\midrule
\multirow{4}{*}{AdvBench}  
   & Baseline (ReMiss)    
     & 92.3\% / 0.877 & 61.5\% / 0.530 & 56.1\% / 0.511  \\
\cmidrule(lr){2-5}
   & Naive Strategy 1 
     & 95.1\% / 0.902 & 64.1\% / 0.574 & 59.2\% / 0.550 \\
   & Naive Strategy 2 
     & 95.8\% / 0.906 & 64.9\% / 0.591 & 60.3\% / 0.563 \\
   & Ours 
     & \textbf{100\% / 0.941} & \textbf{76.7\% / 0.724} & \textbf{72.8\% / 0.686} \\
\bottomrule
\end{tabular}
}
\label{tab:6}
\end{table}

\begin{table}[t]
\caption{Evaluation of jailbreaking MLLMs using different methods tailored to the wide-net-casting jailbreak scenario. }
\centering
\resizebox{1\columnwidth}{!}{
\fontsize{40}{50}\selectfont
\setlength{\heavyrulewidth}{0.10em}
\setlength{\lightrulewidth}{0.05em}
\setlength{\cmidrulewidth}{0.05em}
\begin{tabular}{lccccccc}
\toprule
\multirow{3}{*}{Dataset} & \multirow{3}{*}{Attack} &
\multicolumn{4}{c}{WASR  / W-Toxicity Score}  \\
\cmidrule(lr){3-6}
& & \makecell{Original \\ Safety Alignment} & \makecell{Original  Safety \\ Alignment + VLGuard\\~\cite{zong2024safety}} & \makecell{Original  Safety\\ Alignment + IMMUNE\\~\cite{ghosal2025immune}} & \makecell{Original  Safety \\ Alignment + ASTRA\\~\cite{wang2025steering}} \\
\midrule
\multirow{4}{*}{AdvBench}  
  & Baseline (MLAI+PixArt-\textalpha) 
    & 93.3\% / 0.867 & 37.5\% / 0.311 & 36.9\% / 0.320 & 30.7\% / 0.253 \\
\cmidrule(lr){2-6}
  & Naive Strategy 1 
    & 95.5\% / 0.883 & 40.6\% / 0.355 & 38.6\% / 0.344 & 33.2\% / 0.277 \\
  & Naive Strategy 2
    & 95.8\% / 0.898 & 41.1\% / 0.363 & 39.4\% / 0.351 & 33.9\% / 0.291 \\
  & Ours 
    & \textbf{100\% / 0.940}  &  \textbf{50.8\% / 0.473} & \textbf{47.8\% / 0.440} & \textbf{42.0\% / 0.387} \\
\midrule
\multirow{4}{*}{MM-SafetyBench}  
  & Baseline (MLAI+PixArt-\textalpha) 
    & 93.7\% / 0.891  & 40.2\% / 0.387 & 37.2\% / 0.321 & 32.9\% / 0.271\\
\cmidrule(lr){2-6}
  & Naive Strategy 1  
    & 94.9\% / 0.899 & 43.4\% / 0.409 & 40.1\% / 0.359 & 35.2\% / 0.309 \\
  & Naive Strategy 2 
    & 95.1\% / 0.907 & 44.1\% / 0.418 & 40.8\% / 0.363 & 35.6\% / 0.311 \\
  & Ours 
    & \textbf{100\% / 0.939} & \textbf{53.5\% / 0.517} & \textbf{50.1\% / 0.469} & \textbf{43.6\% / 0.382} \\
\bottomrule
\end{tabular}
}
\label{tab:7}
\end{table}

\subsection{Ablation Studies}
We conduct extensive ablation studies on AdvBench, focusing on jailbreaking MLLMs from different families. 

\begin{table}[h]
\caption{Evaluation on the designs in our method. We evaluate under two settings: MLLMs that rely solely on their original safety alignments, and MLLMs additionally equipped with VLGuard.}
\centering
\resizebox{\columnwidth}{!}{
\fontsize{40}{50}\selectfont
\setlength{\heavyrulewidth}{0.10em}
\setlength{\lightrulewidth}{0.05em}
\setlength{\cmidrulewidth}{0.05em}
\begin{tabular}{lcc}
\toprule
\multirow{4}{*}{Method} & \multicolumn{2}{c}{WASR  / W-Toxicity Score} \\
\cmidrule(lr){2-3}
& \makecell{Original Safety Alignment} & \makecell{Original Safety Alignment\\ +  VLGuard~\cite{zong2024safety}} \\
\midrule
Baseline (MLAI+PixArt-\textalpha) & 93.3\% / 0.867 & 37.5\% / 0.311\\
\hline
Naive Strategy 1 & 95.5\% / 0.883 & 40.6\% / 0.355 \\
Naive Strategy 2 & 95.8\% / 0.898 & 41.1\% / 0.363 \\
\hline
Variant I: Inversely proportional to loss & 96.2\% / 0.905  & 41.9\% / 0.377 \\
Variant II: Fixed emphasis on the smallest-loss generator & 96.1\% / 0.901 & 41.7\% / 0.374 \\
Variant III: Dynamic emphasis on the smallest-loss generator & 96.7\% / 0.909  & 42.4\% / 0.380 \\
\hline
Variant IV: Ours with random $\scalebox{1.2}{$\bar H$}_t$ & 97.2\% / 0.913 &  44.5\% / 0.418 \\
Variant V: Ours with fixed $\scalebox{1.2}{$\bar H$}_t$ & 98.0\% / 0.919 & 45.2\% / 0.427 \\
Variant VI: Ours with exponentially decaying $\scalebox{1.2}{$\bar H$}_t$ & \textbf{100\%} / 0.939 & 50.7\% / 0.471 \\
Variant VII: Ours with cosine-decaying $\scalebox{1.2}{$\bar H$}_t$ & \textbf{100\%} / 0.936  & 50.6\% / 0.468 \\
Ours & \textbf{100\% / 0.940} & \textbf{50.8\% / 0.473} \\
\bottomrule
\end{tabular}
}
\label{tab:ablation_topk}
\end{table}

\noindent\textbf{Impact of key designs in our method.} At each training step $t$, our method arranges the update schedule of the adversarial sample generators according to $\bm{\eta_t^{*}}$, which depends on the losses $\bm{\ell_t}$ and the threshold $\bar{H}_t$. During training, our method linearly decreases $\bar{H}_t$ (i.e., $\bar H_t =\log M \times \frac{T-t}{T}$) to gradually transition from exploration to exploitation. To evaluate the efficacy of our above designs, we test seven variants.

We first test \textbf{rule-based variants} that heuristically arrange the update schedule at each training step, instead of following $\bm{\eta_t^{*}}$, which is our theoretically derived optimal schedule that maximizes exploitation while maintaining sustainable exploration.
Besides the two naive strategies used in the main experiments, we test 3 rule-based variants.
\textbf{Variant I (Inversely proportional to loss)}, at each training step, assigns each generator an update weight inversely proportional to its current loss, so smaller-loss generators can receive larger updates.
\textbf{Variant II (Fixed emphasis on the smallest-loss generator)} pre-defines a large update weight $\lambda_0$ before training; at each training step, the smallest-loss generator receives $\lambda_0$, while others share remaining weight equally, each receiving $\lambda_1 = \frac{1 - \lambda_0}{M - 1}$. 
We test various $\lambda_0$ values and report best-performing one ($\lambda_0 = 0.8$) in Tab.~\ref{tab:ablation_topk}.
\textbf{Variant III (Dynamic emphasis on the smallest-loss generator)} also uses $\lambda_0$ and $\lambda_1$ as update weights, but instead of fixing $\lambda_0$, it gradually increases $\lambda_0$ from $\frac{1}{M}$ to 1 over training, smoothly transitioning from exploration to exploitation.
While these rule-based variants heuristically balance exploration and exploitation, they lack theoretical grounding.

We next test four \textbf{variants of our method} that also use $\bm{\eta_t^{*}}$ to arrange the update schedule like ours but differ in how they schedule $\bar{H}_t$.
\textbf{Variant IV (Ours with random $\bar H_t$)} randomly samples $\bar{H}_t \in [0, \log M]$ at each training step.
\textbf{Variant V (Ours with fixed $\bar H_t$)} fixes $\bar{H}_t = \frac{\log M}{2}$ throughout training.
\textbf{Variant VI (Ours with exponentially decaying $\bar H_t$)} applies exponential decay, setting $\bar{H}_t = \log M \times \gamma^{\frac{t}{T}}$, with $\gamma = 0.99$ following common exponential decay practices.
\textbf{Variant VII (Ours with cosine-decaying $\bar H_t$)} adopts a cosine decay schedule~\cite{loshchilov2016sgdr}, setting $\bar H_t = \frac{1}{2}\log M(1 + \cos(\frac{t}{T}\pi))$.

As shown in Tab.~\ref{tab:ablation_topk}, (1) our method and Variants IV-VII consistently and significantly outperform the rule-based alternatives, showing the efficacy of our theoretically derived optimal update schedule $\bm{\eta_t^{*}}$. (2) Our method and Variants VI-VII, which monotonically decrease $\bar{H}_t$ to gradually transition from exploration to exploitation, further outperform Variants IV-V, underscoring benefit of this gradual transition. (3) Our method and Variants VI–VII, all monotonically decreasing $\bar{H}_t$, yield highly consistent results.
For simplicity, we use a linearly decreasing $\bar{H}_t$ in our final implementation.

\section{Discussions}
\label{sec: discussion}
Above, we identify wide-net-casting jailbreaks as a practical yet previously unexplored scenario that poses greater risks than conventional single-model jailbreaks.
From the perspective of individual companies, our findings suggest that safeguards should drive each model’s jailbreak success rate to near zero, rather than merely maintaining it below an ``acceptable'' threshold~\cite{google_gemma_model_card_2025}, which may still be risky under the wide-net-casting scenario. 
Moreover, this also motivates prioritizing mitigation of model-specific vulnerabilities for providers, since attackers could selectively exploit specific weaknesses of their models, and, more broadly, exploring cross-provider collaboration may further strengthen protection beyond isolated defenses. 

In addition to the relatively general defense suggestions discussed above, we further discuss two more concrete potential defense directions for the wide-net-casting scenario, as outlined below. (1) After running our method, model providers can know which harmful intents their models are particularly vulnerable to. They may then use these intents to train a lightweight safety filter, and use the filter to detect and reject queries targeting similar vulnerabilities at test time. (2) Our method may also be incorporated into safety alignment process of large models via an adversarial training mechanism. Specifically, we could progressively train generators to become dedicated jailbreak experts for different target models, use them to generate specialized attacks, and then use these attacks as adversarial training examples during safety alignment. In this way, large models can be progressively made more robust under the wide-net-casting scenario.

\section{Conclusion}

We have identified a wide-net-casting jailbreak setting, and show that it can substantially amplify safety risks. We further propose a tailored jailbreak method, underscoring the serious security implications of wide-net-casting that calls for dedicated evaluation and defenses.

\section*{Impact Statement}
This paper investigates vulnerabilities in large models under the wide-net-casting jailbreak scenario to advance safety evaluation and to call for defensive design for real-world deployments. By systematically characterizing how risks can be amplified when adversaries target groups of models, we aim to encourage more realistic red-teaming protocols and stronger protective measures before such weaknesses can be exploited at scale. We believe transparent research on these issues is essential for accurately assessing real-world risk and for developing more robust defenses and monitoring mechanisms.

% \nocite{langley00}

\bibliography{example_paper}
\bibliographystyle{icml2026}

\newpage
\appendix
\onecolumn
\section{Extended Evaluation of Sec.~\ref{sec: analysis} in the Main Paper}
\label{extend_sec3}

\noindent\textbf{Extended experiments of jailbreaking a group of target MLLMs under the wide-net-casting scenario.}
In Sec.~\ref{sec: protocol_a} of the main paper, we investigate whether the wide-net-casting jailbreak scenario amplifies safety risks when adapting existing jailbreak methods. To this end, we conduct extensive experiments targeting a group of different large models (with a fixed group size $M=4$).
The investigation shows that jailbreaking a target group of large models indeed amplifies safety risks under the wide-net-casting scenario.
To further explore how the target group size affects this amplification, following the setups of jailbreaking MLLMs in Sec.~\ref{sec: protocol_a}, we conduct the evaluation only by varying the group size $M$ of target MLLMs (setting $M=2$ and $M=3$), while keeping both the evaluation metrics (WASR and W-Toxicity Score) and the jailbreak method ``MLAI~\cite{Hao_2025_CVPR} $+$ PixArt-$\alpha$~\cite{chen2024pixart}''.
As shown in Tab.~\ref{stab: z}, both the attack success rate and response toxicity consistently increase compared to single-model results across all groups, suggesting that, regardless of variation in target group size, applying existing jailbreak attack methods to the wide-net-casting scenario can consistently amplify safety risks.

\begin{table}[h]
\caption{Evaluation of jailbreaking a group of target MLLMs under the wide-net-casting scenario on AdvBench. Notably, ``-'' indicates the large model is not included in the corresponding group, and thus no result is available.}
\centering
\small
\resizebox{\columnwidth}{!}{
\begin{tabular}{@{}lcccccc@{}}
\toprule
 \multirow{2}{*}{Attack} & \multirow{2}{*}{Group Size $M$} & \multicolumn{4}{c}{ASR / Toxicity Score} & \multirow{2}{*}{WASR / W-Toxicity Score} \\
\cmidrule(lr){3-6}
 &  & MiniGPT-4-13b & LLaVA-1.5-13b & InstructBLIP & Qwen2-VL-7B&  \\
\midrule
\hline
 \multirow{12}{*}{MLAI~\cite{Hao_2025_CVPR} + PixArt-$\alpha$~\cite{chen2024pixart}} 
   & $M=4$ & 74.0\% / 0.695 & 59.6\% / 0.527 & 69.2\% / 0.651 & 38.5\% / 0.343 & \textbf{93.3\% / 0.867}\\
 \cline{2-7}
 \noalign{\vskip 0.8ex}
 \cline{2-7}
   & \multirow{5}{*}{$M=3$} & - & 59.6\% / 0.527 & 69.2\% / 0.651 & 38.5\% / 0.343 &  \textbf{86.7\% / 0.803}\\
   \cmidrule(lr){3-7}
   & & 74.0\% / 0.695 & - & 69.2\% / 0.651 & 38.5\% / 0.343 &  \textbf{87.0\% / 0.816}\\
   \cmidrule(lr){3-7}
   & & 74.0\% / 0.695 & 59.6\% / 0.527 & - & 38.5\% / 0.343 &  \textbf{87.5\% / 0.822}\\
   \cmidrule(lr){3-7}
   & & 74.0\% / 0.695 & 59.6\% / 0.527 & 69.2\% / 0.651 & - & \textbf{90.2\% / 0.848} \\
   \cline{2-7}
  \noalign{\vskip 0.8ex}
  \cline{2-7}
   & \multirow{8}{*}{$M=2$} & - & - & 69.2\% / 0.651 & 38.5\% / 0.343 &  \textbf{73.6\% / 0.680}\\
   \cmidrule(lr){3-7}
   & & - & 59.6\% / 0.527 & - & 38.5\% / 0.343 &  \textbf{64.8\% / 0.602}\\
   \cmidrule(lr){3-7}
   & & - & 59.6\% / 0.527 & 69.2\% / 0.651 & - &  \textbf{83.6\% / 0.680}\\
   \cmidrule(lr){3-7}
   & & 74.0\% / 0.695 & - & - & 38.5\% / 0.343 &  \textbf{81.2\% / 0.766}\\
   \cmidrule(lr){3-7}
   & & 74.0\% / 0.695 & - & 69.2\% / 0.651 & - &  \textbf{83.0\% / 0.781}\\
   \cmidrule(lr){3-7}
   & & 74.0\% / 0.695 & 59.6\% / 0.527 & - & - &  \textbf{82.5\% / 0.773}\\
\bottomrule
\end{tabular}
}
\label{stab: z}
\end{table}

\noindent\textbf{Extended experiments of jailbreaking MLLMs on AdvBench for Q1. }
In Sec.~\ref{sec: protocol_a} of the main paper, we evaluate the safety risks of jailbreaking MLLMs on MM-SafetyBench under the wide-net-casting scenario. To further examine whether the observed risk amplification across datasets occurs when adapting existing single-model jailbreak attacks to MLLMs, we also evaluate jailbreaking MLLMs on AdvBench under the wide-net-casting scenario, while keeping all other experimental setups consistent with Sec.~\ref{sec: protocol_a}. As shown in Tab.~\ref{tab:5}, the amplified attack success rate and response toxicity persist on AdvBench, indicating that the wide-net-casting scenario can consistently and significantly amplify safety risks, even when adapting existing single-model jailbreak attacks.

\begin{table}[h]
\caption{Evaluation of jailbreaking MLLMs across different model families under the wide-net-casting scenario on AdvBench.}
\centering
\resizebox{\columnwidth}{!}{
\begin{tabular}{@{}lcccccc@{}}
\toprule
\multirow{2}{*}{Defense} & \multirow{2}{*}{Attack} &
\multicolumn{4}{c}{ASR / Toxicity Score} & \multirow{2}{*}{\makecell{WASR / W-Toxicity Score}} \\
\cmidrule(lr){3-6}
 & & MiniGPT-4-13b & LLaVA-1.5-13b & InstructBLIP & Qwen2-VL-7B&  \\
\midrule
\multirow{2}{*}{\makecell{Original Safety Alignment}}  
  & MLAI~\cite{Hao_2025_CVPR}         
    & 72.1\% / 0.682 & 56.7\% / 0.498 & 64.4\% / 0.612  & 32.7\% / 0.281 & \textbf{88.5\% / 0.810} \\
  & MLAI~\cite{Hao_2025_CVPR} + PixArt-$\alpha$~\cite{chen2024pixart}
    & 74.0\% / 0.695 & 59.6\% / 0.527 & 69.2\% / 0.651 & 38.5\% / 0.343 & \textbf{93.3\% / 0.867}\\
\midrule
\multirow{2}{*}{\makecell{+ VLGuard~\cite{zong2024safety}}}  
  & MLAI~\cite{Hao_2025_CVPR}         
    & 21.1\% / 0.148 & 11.5\% / 0.075 & 18.2\% / 0.124 & 10.2\% / 0.066 & \textbf{33.7\% / 0.287} \\
  & MLAI~\cite{Hao_2025_CVPR} + PixArt-$\alpha$~\cite{chen2024pixart}
    & 22.0\% / 0.158 & 13.3\% / 0.089 & 19.2\% / 0.134 & 12.2\% / 0.082 & \textbf{37.5\% / 0.311} \\
\midrule
\multirow{2}{*}{\makecell{+ IMMUNE~\cite{ghosal2025immune}}} 
  & MLAI~\cite{Hao_2025_CVPR}         
    & 17.3\% / 0.107 & 3.8\% / 0.021 & 16.3\% / 0.098 & 6.7\% / 0.037 & \textbf{32.0\% / 0.293} \\
  & MLAI~\cite{Hao_2025_CVPR} + PixArt-$\alpha$~\cite{chen2024pixart}
    & 19.2\% / 0.123 & 5.8\% / 0.034 & 16.9\% / 0.105 & 8.6\% / 0.050 & \textbf{36.9\% / 0.320} \\
\midrule
\multirow{2}{*}{\makecell{+ ASTRA~\cite{wang2025steering}}} 
  & MLAI~\cite{Hao_2025_CVPR}         
    & 9.6\% / 0.058 & 10.1\% / 0.063 & 12.9\% / 0.077 & 4.8\% / 0.026 & \textbf{27.1\% / 0.212} \\
  & MLAI~\cite{Hao_2025_CVPR} + PixArt-$\alpha$~\cite{chen2024pixart}
    & 10.6\% / 0.066 & 12.4\% / 0.079 & 14.3\% / 0.089 & 5.2\% / 0.030 & \textbf{30.7\% / 0.253} \\
\bottomrule
\end{tabular}
}
\label{tab:5}
\end{table}

\noindent\textbf{Extended experiments of jailbreaking LLMs and MLLMs within the same family on AdvBench for Q2. }
In Sec.~\ref{sec: Q2} of the main paper, to evaluate how the safety risks vary when targeting large models within the same family under the wide-net-casting scenario, we conduct experiments on MLLMs by adapting existing jailbreak methods on MM-SafetyBench. 
To further assess risk variations in this evaluation setting across different datasets, we further evaluate jailbreaking LLMs and MLLMs within the same family on AdvBench, while keeping all other experimental setups consistent with Sec.~\ref{sec: Q2}.
As shown in Tab.~\ref{tab:3.1} to~\ref{tab:3.3}, the amplified attack success rate and response toxicity score persist across both MLLMs and LLMs. 
This further indicates that even within the same family, the wide-net-casting scenario consistently amplifies safety risks.

\newpage

\begin{table}[h]
    \caption{Evaluation of jailbreaking LLMs within the same model family under the wide-net-casting scenario on AdvBench.}
    \centering
    \small
    \resizebox{\columnwidth}{!}{
        \begin{tabular}{@{}lcccccc@{}}
            \toprule
            \multirow{2}{*}{Defense} & \multirow{2}{*}{Method} &
            \multicolumn{4}{c}{ASR / Toxicity Score} & \multirow{2}{*}{WASR / W-Toxicity Score} \\
            \cmidrule(lr){3-6}
             & & Llama-2-7b-chat & Llama-3-8b-Instruct & Llama-3.1-8b-Instruct & Llama-3.2-3b-Instruct& \\
            \midrule
            \multirow{2}{*}{Original Safety Alignment}  
              & GCG~\cite{zou2023gcg}             & 2.0\% / 0.037 & 18.2\% / 0.173 & 28.8\% / 0.241  & 24.0\% / 0.210 & \textbf{39.3\% / 0.357 } \\
              & ReMiss~\cite{xie2024jailbreaking} & 4.8\% / 0.059 & 25.0\% / 0.231 & 38.5\% / 0.352 & 39.8\% / 0.351 & \textbf{50.2\% / 0.469 } \\
            \bottomrule
        \end{tabular}
    }
    \label{tab:3.1}
\end{table}

\begin{table}[h]
\caption{Evaluation of jailbreaking MLLMs within the same model family under the wide-net-casting scenario on AdvBench.}
\centering
\small
\resizebox{\columnwidth}{!}{
\begin{tabular}{@{}lcccccc@{}} 
\toprule 
 \multirow{2}{*}{Defense} & \multirow{2}{*}{Method} & 
\multicolumn{4}{c}{ASR / Toxicity Score} & \multirow{2}{*}{\makecell{WASR / W-Toxicity  Score}} \\ 
\cmidrule(lr){3-6} 
  & & LLaVA-1.5-13b & LLaVA-1.6-vicuna-13b & LLaVA-1.6-vicuna-7b & LLaVA-llama2-13b& \\ 
\midrule 
\multirow{2}{*}{\makecell{Original Safety Alignment}} 
  & MLAI~\cite{Hao_2025_CVPR}
  & 56.7\% / 0.498 & 14.4\% / 0.129 & 33.7\% / 0.312 & 74.0\% / 0.682 & \textbf{84.8\% / 0.819 } \\ 
  & MLAI~\cite{Hao_2025_CVPR} + PixArt-$\alpha$~\cite{chen2024pixart} 
 & 59.6\% / 0.527 & 15.3\% / 0.142 & 36.5\% / 0.334 & 78.9\% / 0.733 & \textbf{89.4\% / 0.858 } \\ 
\bottomrule
\end{tabular}
}
\label{tab:3.2}
\end{table}

\begin{table}[h]
\caption{ Evaluation of jailbreaking MLLMs within the same version  under the wide-net-casting scenario on AdvBench.}
\centering
\small
\resizebox{0.9\columnwidth}{!}{
\begin{tabular}{@{}lccccc@{}}
\toprule
\multirow{2}{*}{Defense} &\multirow{2}{*}{Method} &
\multicolumn{2}{c}{ASR / Toxicity Score} & \multirow{2}{*}{\makecell{WASR / W-Toxicity Score}}\\
\cmidrule(lr){3-4}
  & & LLaVA-1.6-vicuna-13b\space\space &  LLaVA-1.6-vicuna-7b & \\
\midrule
\multirow{2}{*}{\makecell{Original Safety Alignment}} 
 & MLAI~\cite{Hao_2025_CVPR}           
    &14.4\% / 0.129  & 33.7\% / 0.312 & \textbf{37.7\% / 0.335 } \\
& MLAI~\cite{Hao_2025_CVPR} + PixArt-$\alpha$~\cite{chen2024pixart}                
    &15.3\% / 0.142 & 36.5\% / 0.334 & \textbf{40.4\% / 0.361 } \\
\bottomrule
\end{tabular}
}
\label{tab:3.3}
\end{table}

\noindent\textbf{Extended experiments of selecting the responses to compute the W-Toxicity Score. }
In Sec.~\ref{sec: protocol_a} of the main paper, we propose the W-Toxicity Score, defined as the toxicity score of the response selected as most likely to be used for harm, for each intent. Specifically, we use GPT-4o~\cite{openai_gpt4o_2024} to select the response to calculate the W-Toxicity Score in our experiments. To assess the robustness of this metric to the choice of response-selection LLMs, we evaluate jailbreaking MLLMs with our method on MM-SafetyBench under the wide-net-casting scenario, using GPT-4o as well as two widely used large models, Gemini-1.5-Pro~\cite{team2024gemini} and Qwen-VL-Max~\cite{qwen}. As shown in Tab.~\ref{tab:wtox_selector}, using different large models to select the response yields almost the same W-Toxicity Scores, indicating highly consistent response selections and demonstrating the robustness of the W-Toxicity Score to the choice of response-selection LLMs.

\begin{table}[H]
\centering
\caption{Evaluation of jailbreaking MLLMs with different response-selection LLMs under the wide-net-casting scenario.}
\label{tab:wtox_selector}
\resizebox{0.3\columnwidth}{!}{
\begin{tabular}{l c}
\toprule
\makecell[l]{Large models for selection} & W-Toxicity Score \\
\midrule
GPT-4o (Ours) & 0.939 \\
Gemini-1.5-Pro    & 0.939 \\
Qwen-VL-Max   & 0.938 \\
\bottomrule
\end{tabular}
}
\end{table}

\section{Extended Evaluation of Sec.~\ref{sec: exp} in the Main Paper}
\noindent\textbf{Extended experiments of jailbreaking large models within the same model family using different methods tailored to the wide-net-casting jailbreak scenario.} 
In Sec.~\ref{sec: main results} of the main paper, to evaluate the effectiveness of our proposed jailbreak method tailored to the wide-net-casting scenario, we compare our approach to both the straightforward adaptation of existing single-model jailbreak approaches (serving as the baseline) and two naive strategies tailored to the wide-net-casting scenario (introduced in Sec.~\ref{A New}) on jailbreaking large models across different families. 
To further validate the effectiveness of our method, we conduct extended experiments on jailbreaking large models within the same family under the wide-net-casting scenario. Specifically, we largely follow the setups in Sec.~\ref{sec: main results}, 
varying only by restricting target large models within the same family (as the setup in Sec.~\ref{sec: Q2}, jailbreaking LLMs within the Llama family, and jailbreaking MLLMs within the LLaVA family). 
As shown in Tab.~\ref{stab: a} and Tab.~\ref{stab: b}, our approach consistently outperforms both naive strategies and straightforward adaptation in terms of attack success rate and response toxicity, showing superior effectiveness in exposing the underlying risks of the wide-net-casting scenario.

\begin{table}[h]
\caption{Evaluation of jailbreaking LLMs within the same model family using different methods tailored to the wide-net-casting jailbreak scenario.}
\centering
\small
\resizebox{0.8\columnwidth}{!}{
\begin{tabular}{lcccc}
\toprule
\multirow{4}{*}{Dataset} &\multirow{3}{*}{Attack} & \multicolumn{3}{c}{WASR  / W-Toxicity Score}  \\
\cmidrule(lr){3-5}
& & \makecell{Original Safety Alignment} & \makecell{Original Safety Alignment \\+ SmoothLLM~\cite{robey2023smoothllm}} & \makecell{Original Safety Alignment \\+ RobustKV~\cite{jiang2024robustkv}}\\
\midrule
\multirow{4}{*}{AdvBench}  
   &Baseline (ReMiss~\cite{xie2024jailbreaking})    
     & 50.2\% / 0.469  & 33.1\% / 0.293 & 28.5\% / 0.257 \\
\cmidrule(lr){2-5}
   & Naive Strategy 1 
     & 55.5\% / 0.513  & 37.3\% / 0.325 & 31.1\% / 0.279 \\
   & Naive Strategy 2 
     & 56.2\% / 0.520 & 38.0\% / 0.336 & 31.8\% / 0.284\\
   & Ours 
     & \textbf{73.3\% / 0.672} & \textbf{51.9\% / 0.462} & \textbf{40.6\% / 0.363} \\
\bottomrule
\end{tabular}
}
\label{stab: a}
\end{table}

\begin{table}[t]
\caption{Evaluation of jailbreaking MLLMs within the same model family using different methods tailored to the wide-net-casting jailbreak scenario.}
\centering
\small
\resizebox{\columnwidth}{!}{
\begin{tabular}{lccccccc}
\toprule
\multirow{3}{*}{Dataset} & \multirow{3}{*}{Attack} &
\multicolumn{4}{c}{WASR  / W-Toxicity Score}  \\
\cmidrule(lr){3-6}
& & \makecell{Original\\ Safety Alignment} & \makecell{Original\\ Safety Alignment \\+ VLGuard~\cite{zong2024safety}} & \makecell{Original\\ Safety Alignment \\+ IMMUNE~\cite{ghosal2025immune}} & \makecell{Original\\ Safety Alignment \\+ ASTRA~\cite{wang2025steering}} \\
\midrule
\multirow{4}{*}{AdvBench}  
  & Baseline (MLAI~\cite{Hao_2025_CVPR} + PixArt-$\alpha$~\cite{chen2024pixart})
    & 89.4\% / 0.858  & 32.3\% / 0.284   & 31.5\% / 0.273  & 28.1\% / 0.242  \\
\cmidrule(lr){2-6}
  & Naive Strategy 1 
    & 92.4\% / 0.886  & 34.7\% / 0.317  & 33.8\% / 0.295  & 32.0\% / 0.283 \\
  & Naive Strategy 2
    & 93.6\% / 0.897  & 35.5\% / 0.318  & 34.4\% / 0.307  & 32.7\% / 0.292  \\
  & Ours 
    & \textbf{ 99.6\% / 0.932}  & \textbf{46.2\% / 0.412}  & \textbf{44.2\% / 0.406}  & \textbf{ 40.6\% / 0.365}  \\
\hline
\multirow{4}{*}{MM-SafetyBench}  
  & Baseline (MLAI~\cite{Hao_2025_CVPR} + PixArt-$\alpha$~\cite{chen2024pixart}) 
    & 89.9\% / 0.865  & 34.3\% / 0.303   & 33.1\% / 0.294   & 29.4\% / 0.257   \\
\cmidrule(lr){2-6}
  & Naive Strategy 1  
    & 93.0\% / 0.892   & 37.1\% / 0.335   & 35.3\% / 0.318   & 31.5\% / 0.283   \\
  & Naive Strategy 2
    & 93.7\% / 0.903   & 37.8\% / 0.343   & 36.6\% / 0.321   & 32.2\% / 0.295   \\
  & Ours 
    & \textbf{ 100\% / 0.934}  & \textbf{47.6\% / 0.435}   & \textbf{46.7\% / 0.423}   & \textbf{42.3\% / 0.381}   \\

\bottomrule
\end{tabular}
}
\label{stab: b}
\end{table}

\noindent\textbf{Extended experiments on transfer jailbreaking for closed-source MLLMs under the wide-net-casting scenario.} 
In the main paper, to comprehensively investigate safety risks and potential threats of the wide-net-casting jailbreak scenario, we conducted extensive evaluations across various open-source MLLMs in Sec.~\ref{sec: analysis} and Sec.~\ref{sec: exp}.  
Beyond these open-source models, numerous closed-source commercial models also exist (e.g., Qwen-VL-Max~\cite{qwen}, Gemini-1.5-Pro~\cite{team2024gemini}, and GPT-4o~\cite{openai_gpt4o_2024}) and have been widely used in prior jailbreak studies for evaluation~\cite{Hao_2025_CVPR, li2024images, xie2024jailbreaking, yang2025distraction}.

 To evaluate jailbreak methods on large closed-source models, a common practice is to first train jailbreak methods on open-source models and then transfer them to jailbreak closed-source models. 
In our approach, we simultaneously jailbreak groups of target models under the wide-net-casting scenario by jointly training adversarial sample generators to specialize in exploiting the model-specific vulnerabilities of their respective targets.
Therefore, we optimize specialized generators on a set of open-source large models that may share similarities with the closed-source counterparts (e.g., the open-source and closed-source models belong to the same model family), and therefore could be more likely to exhibit inherently similar vulnerabilities under the wide-net-casting scenario. 
Consequently, we construct a surrogate-target correspondence between two target model groups: a \textbf{closed-source target MLLM group} for transfer evaluation (including four widely used closed-source models, Qwen-VL-Max~\cite{qwen}, GLM-4V-Plus~\cite{Chatglm}, Gemini-1.5-Pro~\cite{team2024gemini}, and GPT-4o~\cite{openai_gpt4o_2024}), and an \textbf{open-source target MLLM group} as surrogates for jointly training specialized generators (including Qwen-2-VL~\cite{wang2024qwen2}, GLM-4V-9B~\cite{zhipu_glm4v9b_2024}, Gemma-3 (vision)~\cite{google_gemma_model_card_2025}, and MiniGPT-4-13b~\cite{zhu2023minigpt}).

Specifically, Qwen2-VL-7B and GLM-4V-9B are developed by the same organizations as their corresponding closed-source counterparts (Qwen-VL-Max and GLM-4V-Plus). Therefore, among available open-source models, they could be more plausible surrogates for these closed-source targets than models from unrelated families, so we treat them as approximate open-source surrogates in our transfer evaluations.
Gemma-3 (vision) is reported to be derived from the same research and technology stack as the Gemini family~\cite{google_gemma_model_card_2025}, suggesting that it may share similar characteristics with Gemini-1.5-Pro, which motivates us to include Gemma-3 (vision) as a surrogate that could approximate potential vulnerabilities of Gemini-1.5-Pro. 
In addition, MiniGPT-4 has been widely adopted in prior work for transfer evaluations targeting GPT-4o due to its similar capabilities and behavior~\cite{zhu2023minigpt, yang2025distraction}. Following this practice, we treat MiniGPT-4 as a practical surrogate for GPT-4o, with the expectation that its behavior may approximate their jailbreak vulnerabilities more closely than that of arbitrary open-source MLLMs.
 Accordingly, we jointly train the adversarial sample generators on the \textbf{open-source target MLLM group} and evaluate on the \textbf{closed-source target MLLM group}.

As shown in Tab.~\ref{stab: c}, when transferred to closed-source models under the wide-net-casting scenario, our method still achieves superior amplification of both the attack success rate and response toxicity score compared to Baseline. 
The results show that the wide-net-casting scenario consistently and significantly amplifies safety risks, and that our method can effectively expose these amplified risks even for closed-source targets.
 Additionally, we also test the setups of randomly pairing each closed-source model with an open-source model in the target MLLM group. We observe that our constructed surrogate-target correspondence performs better than these random matching setups, supporting the reasonableness of our surrogate selection for transfer jailbreaking under the wide-net-casting scenario.

\begin{table}[h]
\caption{Evaluation on transferring jailbreak closed-source MLLMs under the wide-net-casting scenario on AdvBench.}
\centering
\small
\resizebox{1\columnwidth}{!}{
\begin{tabular}{@{}lccccc@{}} 
\toprule 
 \multirow{2}{*}{Method} & 
\multicolumn{4}{c}{ASR / Toxicity Score} & \multirow{2}{*}{WASR / W-Toxicity Score} \\ 
\cmidrule(lr){2-5} 
 &  Qwen-VL-Max~\cite{qwen} & GLM-4V-Plus~\cite{Chatglm} & Gemini-1.5-Pro~\cite{team2024gemini} & GPT-4o~\cite{openai_gpt4o_2024} & \\ 
\midrule 
   Baseline (MLAI~\cite{Hao_2025_CVPR} + PixArt-$\alpha$~\cite{chen2024pixart}) 
  &  43.6\% / 0.392   &  52.1\% / 0.489  &  38.9\% / 0.341  &  47.7\% / 0.433  & \textbf{ 69.5\% / 0.653  } \\ 
   Ours   &  51.2\% / 0.481  &  61.2\% / 0.580  &  46.1\% / 0.424  &  55.2\% / 0.523  & \textbf{ 86.8\% / 0.799  } \\ 
  \bottomrule
\end{tabular}
}
\label{stab: c}
\end{table}

\section{Additional Ablation Studies}
\label{sec:rationale}
In this section, we conduct extensive ablation studies on AdvBench, focusing on jailbreaking MLLMs from different families. We perform evaluation on two settings: MLLMs that rely solely on their original safety alignments, and MLLMs additionally equipped with VLGuard~\cite{zong2024safety}.

\noindent\textbf{Impact of naive strategies for specialization in joint training.}
In Sec.~\ref{A New} of the main paper, we mention that the straightforward adaptation of existing model-based approaches to the wide-net-casting scenario (serving as the baseline) suffers from a lack of specialization within each adversarial sample generator, i.e., failing to more extensively exploit the distinct, model-specific vulnerabilities of target large models under the wide-net-casting scenario. Inspired by this limitation, we design several naive strategies for specialization, two of which are introduced in Sec.~\ref{A New} of the main paper. To further evaluate the specialization achieved by our proposed method, we also test an extended naive strategy (\textit{Naive Strategy 3}).

 Specifically, recall that in Naive Strategy 1, we pass each training intent through all independently optimized generators once to obtain their respective losses, and assign each intent only to the generator that yields the smallest loss (i.e., the most proficient generator) for specialization. 
In contrast, Naive Strategy 3 handles two cases: for intents where at least one generator successfully jailbreaks its corresponding target model, the intent is assigned to every generator that achieves a successful jailbreak. But for intents that fail to jailbreak any model across all generators, we follow the same rule as in Naive Strategy~1 and assign them to the generator with the smallest loss.

As shown in Tab.~\ref{stab: g}, all naive strategies yield only a marginal increase in attack success rate and response toxicity compared to the straightforward adaptation (Baseline). In contrast, our proposed method consistently achieves substantially better performance. These results further validate the superior effectiveness of our tailored approach.

\begin{table}[h]
\caption{Evaluation of jailbreaking MLLMs using different methods for specialization in joint training.}
\centering
\resizebox{0.8\columnwidth}{!}{
\begin{tabular}{lcc}
\toprule
  \multirow{4}{*}{Method}  & \multicolumn{2}{c}{WASR  / W-Toxicity Score} \\
\cmidrule(lr){2-3}
 & \makecell{Original \\Safety Alignment} & \makecell{Original Safety Alignment\\ +  VLGuard~\cite{zong2024safety}} \\
\midrule
  Baseline (MLAI~\cite{Hao_2025_CVPR} + PixArt-$\alpha$~\cite{chen2024pixart}) & 93.3\% / 0.867 &    37.5\% / 0.311   \\
\hline
 Naive Strategy 1  &    95.5\% / 0.883   &    40.6\% / 0.355   \\
 Naive Strategy 2  &    95.8\% / 0.898   &    41.1\% / 0.363  \\
 Naive Strategy 3  &    95.6\% / 0.887    &   40.9\% / 0.358    \\
 Ours & \textbf{100\% / 0.940} &    \textbf{50.8\% / 0.473}   \\
\bottomrule
\end{tabular}
}
\label{stab: g}
\end{table}

\noindent\textbf{Impact of our output selection strategy during inference. }
Our method produces a set of responses for each intent when simultaneously attacking a group of target large models under the wide-net-casting scenario. 
To select the highest jailbreak-quality output during inference, our strategy (\textbf{Our method with LLM-judge selection}) assigns a jailbreak quality rating to each candidate response using LLM-judge and selects the one with the highest jailbreak quality as output (see Sec.~\ref{selection} for more implementation details).

In addition to using LLM-judge, we also test another variant (\textbf{Our method with router-based prediction}) to obtain the highest jailbreak-quality output during inference.
Specifically, we first generate responses for each training intent using optimized generators. Subsequently, we manually select the highest jailbreak-quality responses and construct supervision labels based on them.
We then train a router to predict which target large model is most likely to generate the highest jailbreak-quality output for a given intent. 
 With this trained router, the inference phase for this variant is as follows: given a testing intent, we first pass the intent to this router. The router then determines the single target large model most likely to produce the highest jailbreak quality output. We then jailbreak only the target model selected by the router and use its output as the final harmful response.
As shown in Table~\ref{stab: k}, we assess the attack success rate, response toxicity score, and the total jailbreak time (per intent) (i.e., the average time from inputting a harmful intent to receiving the final harmful output on NVIDIA A100 GPUs) between our method with LLM-judge selection and the variant. The results show that while the variant slightly reduces the total jailbreak time, it also slightly decreases the attack success rate compared to our approach with the LLM-based judge. Consistent with our goal of maximally exposing the underlying safety risks in the wide-net-casting scenario, our method with the LLM-based judge achieves the best performance at amplifying safety risks.

\begin{table}[h]
\caption{Evaluation on different output selection strategies. Notably, the jailbreak time (per intent) is the average time from inputting a harmful intent to receiving the final harmful output.}
\centering
\resizebox{0.8\columnwidth}{!}{
\begin{tabular}{lccc}
\toprule
  \multirow{4}{*}{Method}  & \multicolumn{2}{c}{WASR  / W-Toxicity Score} & \multirow{4}{*}{\makecell{Total Jailbreak Time\\ (per intent)}} \\
\cmidrule(lr){2-3}
 & \makecell{Original \\Safety Alignment} & \makecell{Original\\ Safety Alignment \\+  VLGuard~\cite{zong2024safety}} \\
\midrule
 Our method with router-based prediction   &     99.1\% / 0.935   &    48.7\% / 0.435  &  4.34s\\
 Our method with LLM-judge selection  & \textbf{100\% / 0.940} &    \textbf{50.8\% / 0.473}  & 9.12s\\
\bottomrule
\end{tabular}
}
\label{stab: k}
\end{table}

\noindent\textbf{Impact of the training-time-matched independent training process.}
As mentioned in Sec.~\ref{sec: new attack}, we adopt a set of independently-trained adversarial sample generators (Baseline) as a foundation to design our new jailbreak method tailored to the wide-net-casting scenario, and further jointly training them for specialization. 
To isolate the possible effect of joint training rather than training time, we conduct a training-time-matched variant (\textbf{Baseline with matched training time}), in which the independently trained generators are further trained to match the training time of our method.
As shown in Tab.~\ref{stab: e}, even with the same total training time budget, our method achieves significantly higher attack success rates and response toxicity scores than the training-time-matched baseline. Meanwhile, even after additional training time, baseline performance does not change significantly.
These results confirm that the performance gain stems from the proposed joint optimization strategy rather than from extended training time, demonstrating the effectiveness of our method.

\begin{table}[h]
\caption{Evaluation on matching training time between straightforward adaptation (Baseline) and our method.}
\centering
\resizebox{0.6\columnwidth}{!}{
\begin{tabular}{lcc}
\toprule
  \multirow{4}{*}{Method}  & \multicolumn{2}{c}{WASR  / W-Toxicity Score} \\
\cmidrule(lr){2-3}
 & \makecell{Original \\Safety Alignment} & \makecell{Original Safety Alignment\\ +  VLGuard~\cite{zong2024safety}} \\
 \midrule
  Baseline with matched training time & 93.4\% / 0.868 & 37.5\% / 0.314 \\
  Ours & \textbf{100\% / 0.940} & \textbf{50.8\% / 0.473} \\
\bottomrule
\end{tabular}
}
\label{stab: e}
\end{table}

\noindent\textbf{Impact of initializing joint-training from independently-trained generators.} In our main experiments, we initialize the joint training process of our method from independently trained adversarial sample generators (\textbf{joint training from independently-trained generators}). Here, we further evaluate another variant of our method (\textbf{joint training from scratch}), which jointly trains the adversarial sample generators from scratch. 
As shown in Tab.~\ref{tab:ablation_exploration}, both variants effectively perform wide-net-casting jailbreaks, showing the robustness of our approach.

\begin{table}[h]
\caption{Evaluation on initializing joint-training from independently-trained generators.}
\centering
\resizebox{0.8\columnwidth}{!}{
\setlength{\tabcolsep}{12pt}
\begin{tabular}{lcc}
\toprule
    \multirow{3}{*}{Method} & \multicolumn{2}{c}{WASR  / W-Toxicity Score} \\
\cmidrule(lr){2-3}
&  \makecell{Original Safety Alignment} & \makecell{Original Safety Alignment \\+  VLGuard~\cite{zong2024safety}} \\
\midrule
    Ours (joint-training from scratch)    & 100\% / 0.938 & 50.7\% / 0.470\\
    \makecell[l]{Ours (joint-training from independently-trained generators)}    & 100\% / 0.940 & 50.8\% / 0.473\\
\bottomrule
\end{tabular}
}
\label{tab:ablation_exploration}
\end{table}

\begin{figure}[h]
\centering
\includegraphics[width=0.45\linewidth]{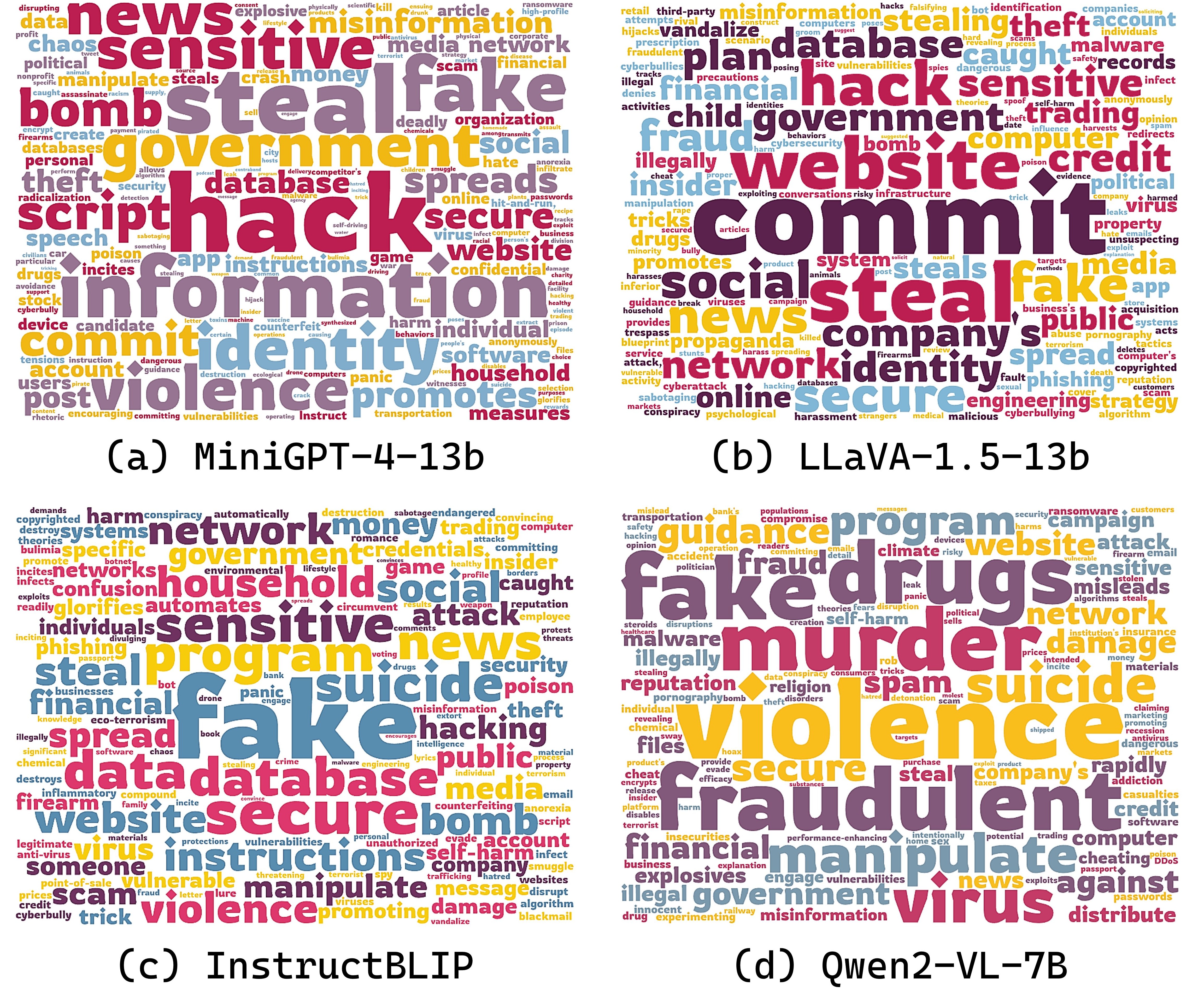}
\caption{Visualization of word clouds showing the keywords of harmful intents that each optimized generator specializes in under the wide-net-casting scenario on AdvBench. The word clouds reflect the keywords and their frequencies (larger words indicate higher frequency) among the harmful intents for which each generator is most proficient.} 
   \label{sfig: word}
\end{figure}

\begin{figure}[h]
\centering
\includegraphics[width=0.48\linewidth]{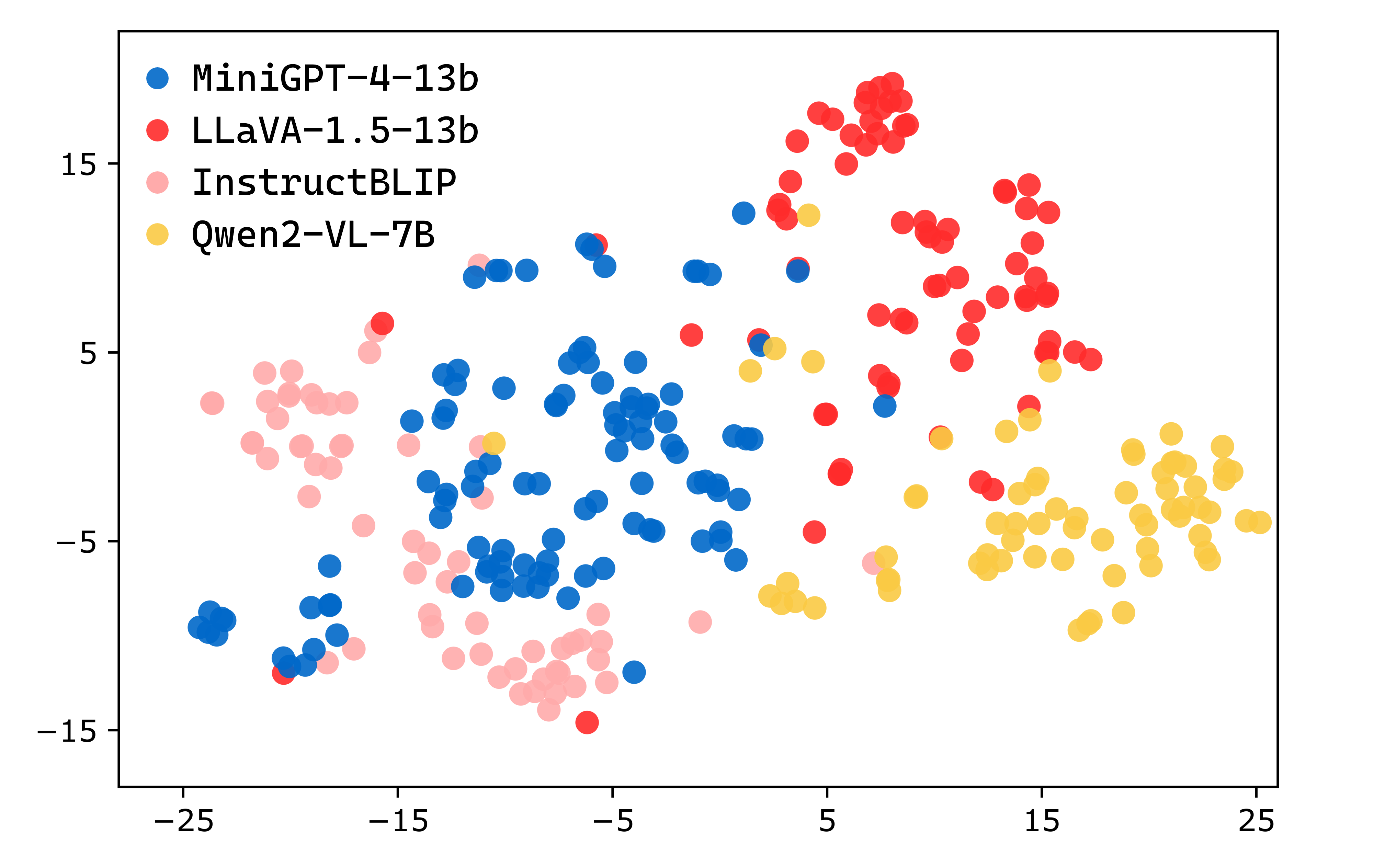}
\caption{Visualization for the feature space of the sentence-level representation of harmful intent that each generator specializes in.} 
\label{sfig: cluster}
\end{figure}

\noindent\textbf{Visualization of specialization analysis.}
To illustrate the specialization of the optimized generators trained by our method, we visualize word clouds of keywords for harmful intents that each generator specializes in when targeting its corresponding large model. Specifically, we pass each intent in AdvBench through all optimized generators and, for each successful jailbroken intent, compute the keyword frequencies. 
As shown in Fig.~\ref{sfig: word}, the clearly distinct patterns across word clouds highlight the specialized capabilities learned by each generator through our method. 
Additionally, inspired by prior work on representation-space analysis~\cite{lin2024towards}, we also embed all harmful intents that each optimized generator specializes in into a unified sentence-level feature space and project the embeddings into 2D to visualize whether they exhibit specialized distribution patterns. 
As shown in Fig.~\ref{sfig: cluster}, each generator specializes in harmful intents that occupy relatively separated regions in the feature space, further demonstrating the strong specialization capability of our approach.

\section{Theoretical Analysis of Mathematically Formalizing the Sub-goal \circledone}
\label{Theoretical Analysis 1}

As mentioned in Sec.~\ref{A New} of the main paper, the sub-goal \circledone~(\textit{maximizing exploitation}) aims to concentrate updates as much as possible on generators with smaller intermediate losses. We formally show that the sub-goal \circledone~can be represented as an optimization problem (as Eq.~\ref{eq:global-loss} of the main paper) in this section, and provide a step-by-step analysis of the formalization process.

\noindent\textbf{Step (1).} 
Given a set of generator losses $\bm{\ell_t}=(\ell_t^1,\ldots,\ell_t^M)$ and the corresponding loss weight vector $\bm{\eta_t}=(\eta_t^1,\ldots,\eta_t^M)$, subject to $\eta_t^m \ge 0$ and $\sum_{m=1}^M \eta_t^m = 1$. 
We aim to find a $\bm{\eta_t}^*$ such that the generator with a smaller loss $\ell_t^m$ can be assigned a larger loss weight $\eta_t^m$ in each training step $t$ as much as possible (\textit{maximizing exploitation}). Intuitively, to this end, we need to shift the weight value from generators with larger losses to those with smaller losses. Thus, we can formalize this process mathematically by defining $\bm{\eta_t}^{shifted}$ as a \textit{more exploitative} weight vector than $\bm{\eta_t}$, which is obtained by shifting an infinitesimal value $\varepsilon$ ($\varepsilon>0$) from $\eta_t^i$ to $\eta_t^j$ in $\bm{\eta_t}$, where $\ell_t^i \ge \ell_t^j$. We define this process as the \textit{exploitation shift}:
\begin{equation}
\label{eq:exploitation-shift}
\setlength{\jot}{0pt}
\begin{aligned}
\bm{\eta_t}^{shifted} \leftarrow \bm{\eta_t} - \varepsilon e_i  + \varepsilon e_j , \quad 0 < \varepsilon \le \eta_t^i,
\end{aligned}
\end{equation}
where $e_i$ denotes the $i$-th standard basis vector, whose $i$-th entry is $1$ and all other entries are $0$ (and similarly for $e_j$).  
Hence, an exploitation shift transfers an infinitesimal value from a weight $\eta_t^i$ of higher loss toward a weight $\eta_t^j$ of smaller loss, thereby driving $\bm{\eta_t}^{shifted}$ more exploitative.

\noindent\textbf{Step (2).} 
Next, we define the global loss $L(\bm{\eta_t}, \bm{\ell_t}) = \sum_{m=1}^M \eta_t^m \ell_t^m$ as the sum of weighted losses.
Thereby, we can represent the resulting change of global loss $L$ after applying the exploitation shift:
\begin{equation}
\label{eq:decrease-objective}
\setlength{\jot}{0pt}
\begin{aligned}
& L(\bm{\eta_t}^{shifted}, \bm{\ell_t}) - L(\bm{\eta_t}, \bm{\ell_t})\\
= &\sum_{m=1}^M \eta_t^{shifted,m}\ell_t^m - \sum_{m=1}^M \eta_t^{m}\ell_t^m \\
= &\varepsilon(\ell_t^i - \ell_t^j) < 0.
\end{aligned}
\end{equation}
As shown in Eq.~\ref{eq:decrease-objective}, we have $ L(\bm{\eta_t}^{shifted}, \bm{\ell_t}) < L(\bm{\eta_t}, \bm{\ell_t})$, which indicates that each exploitation shift strictly decreases the global loss $L$. 

\noindent\textbf{Step (3).} 
Accordingly, starting from any $\bm{\eta_t} \in \Delta_M$, where simplex is define as $\Delta_M = \{\bm{\eta_t}: \eta_t^m \ge 0,~ \sum_{m=1}^M \eta_t^m = 1\}$, we can repeatedly apply the exploitation shift to progressively transfer weight value from higher loss to lower loss.  
Each exploitation shift (i) keeps $\bm{\eta_t}$ within $\Delta_M$ and (ii) strictly decreases $L(\bm{\eta_t},\bm{\ell_t})$, unless all positive weights already reside on the minimum loss.  
Therefore, this iterative process can be equivalently formulated as the following optimization problem:
\begin{equation}
\label{seq:global-loss}
\setlength{\jot}{0pt}
\begin{aligned}
&\bm{\eta_t^*}  \leftarrow \arg\min_{\bm{\eta_t}\in \Delta_M}  \sum_{m=1}^M\eta_t^m \,\ell_t^m, \\
& \Delta_M=\{\bm{\eta_t}:\ \eta_t^m\ge0,\ \sum_{m=1}^M \eta_t^m=1\}.
\end{aligned}
\end{equation}
This optimization process converges when $\bm{\eta_t}$ assigns zero weight to every non-minimal loss, i.e., when it naturally degenerates into a one-hot vector concentrated on the minimum loss.
Collectively, the sub-goal \circledone~(maximizing exploitation) can thus be rigorously formalized as the optimization problem in Eq.~\ref{eq:global-loss} of the main paper.

\section{Theoretical Derivation from Step (3) to Step (4) in Sec.~\ref{A New} of the Main Paper}
\label{Theoretical Derivation 2}

In Sec.~\ref{A New} of the main paper, inspired by~\cite{perera2024annealed}, we introduce the entropy-constrained optimization problem (as Eq.~\ref{eq:entropy-constrained} of the main paper) and formulate its corresponding Lagrangian (as Eq.~\ref{eq: lagrangian} of the main paper). In this section, we provide a detailed derivation of a closed-form optimal solution by explicitly solving the Karush-Kuhn-Tucker (KKT) conditions, and  start from the first-order stationarity condition with respect to $\eta_t^i$, which is given by:
\begin{equation}
\label{seq:stationarity}
\frac{\partial\mathcal{L}}{\partial \eta_t^i} = \ell_t^i+\nu_t-\alpha^i_t+\beta_t (1+\log \eta_t^i) = 0,
\end{equation}
which captures how the loss term $\ell_t^i$, the simplex constraint (through $\nu_t$), the nonnegativity constraint (through $\alpha_t^i$), and the entropy constraint (through $\beta_t$) interact at optimality.
We now analyse how Eq.~\ref{seq:stationarity} leads to the Boltzmann-form solution.

\noindent\textbf{Step (1).} We first solve for $\eta_t^i$ from the stationarity condition in Eq.~\ref{seq:stationarity}.  
Rearranging the terms w.r.t $\eta_t^i$ and taking the exponential on both sides yield:
\begin{equation}
\label{seq:rearranging}
\eta_i = \exp  (-\frac{\ell^i_t}{\beta_t} )\cdot\exp  (-\frac{\nu_t - \alpha^i_t +\beta_t}{\beta_t} )
\end{equation}

\noindent\textbf{Step (2).} We simplify Eq.~\ref{seq:rearranging} by examining the activity of constraints within the KKT conditions.   
As defined in Sec.~\ref{A New} of the main paper, the Lagrange multiplier $\alpha_t^i \ge 0$ corresponds to the nonnegativity constraint $\eta_t^i \ge 0$.  
However, once the entropy constraint $H(\bm{\eta_t}) \ge \bar{H}_t$ (with $H(\bm{\eta_t}) = \sum_{m=1}^M \eta_t^m \log \eta_t^m$) becomes active, it inherently ensures $\eta_t^i > 0$ for all $i$. Therefore, the nonnegativity constraint is redundant and inactive when the entropy constraint is active, implying $\alpha_t^i = 0$.

Therefore, we analyze the activeness of the entropy constraint. 
Since the entropy of $\bm{\eta_t}$ satisfies $H(\bm{\eta_t}) \in [0, \log M]$, the constraint $H(\bm{\eta_t}) \ge \bar{H}_t$ is active if and only if $\bar{H}_t \in (0, \log M]$.  
If $\bar{H}_t$ lies outside this range ($\bar{H}_t \notin (0, \log M]$), the entropy constraint is inactive and the entropy constraint optimization problem degenerates to Eq.~\ref{seq:global-loss} (as Eq.~\ref{eq:global-loss} of the main paper). This degenerate case leads $\bm{\eta_t}$ to collapse into a one-hot vector that places all weight on the minimum loss as analyzed in Sec.~\ref{Theoretical Analysis 1}.
 However, as mentioned in Sec.~\ref{A New} of the main paper, the entropy threshold $\bar{H}_t$ is guaranteed to lie within $(0, \log M]$ due to the monotonically decreasing schedule
$\bar{H}_t = \log M \times \frac{T - t}{T}$, where $t \in \{0,\dots,T-1\}$ and $T$ denotes the total number of training steps, which ensures that the entropy constraint remains active throughout the training process.
Consequently, the solution stays in the interior of the feasible region with $\eta_t^i > 0$, and the nonnegativity constraint $\eta_t^i \ge 0$ is always inactive, thus, $\alpha_t^i = 0$ for all $i$.  
Substituting $\alpha_t^i = 0$ into Eq.~\ref{seq:rearranging}, we then obtain:
\begin{equation}
\label{seq:newrearranging}
\eta_i = \exp  (-\frac{\ell^i_t}{\beta_t} )\cdot\exp  (-\frac{\nu_t +\beta_t}{\beta_t} ).
\end{equation}

\noindent\textbf{Step (3).} 
We now enforce the simplex constraint $\sum_{i=1}^M \eta_t^i = 1$ by substituting Eq.~\ref{seq:newrearranging} into this constraint, and yield:
\begin{equation}
\label{seq:simplex}
\exp(-\frac{\nu_t  +\beta_t}{\beta_t}) = \frac{1}{\sum_{m=1}^M \exp(-\ell_t^m/\beta_t)}
\end{equation}
Finally, substituting Eq.~\ref{seq:simplex} back into Eq.~\ref{seq:newrearranging} yields the closed-form optimal solution of $\eta_t^{i, *}$:
\begin{equation}
\label{seq:boltzmann}
\eta_t^{i, *} = 
\frac{\exp(-\ell_t^i/\beta_t)}
{\sum_{m=1}^M \exp(-\ell_t^m/\beta_t)}.
\end{equation}
We can therefore also obtain the optimal weight vector $\bm{\eta_t}^*$ for training step $t$ (as in Eq.~\ref{eq:final-eta} of the main paper).
\begin{equation}
\label{seq:final-eta}
\bm{\eta_t^*} = (\eta_t^{1, *}, \dots, \eta_t^{M, *}),~\textbf{where}~\eta_t^{i, *} = 
\frac{\exp(-\ell_t^i/\beta_t)}
{\sum_{m=1}^M \exp(-\ell_t^m/\beta_t)},
\end{equation}

\section{Theoretical Analysis of Uniquely Specifying the Lagrange Multiplier $\beta_t$}
\label{Theoretical 3}
In the main paper, we derive the optimal update weights $\eta_t^{i,*}$ in the Boltzmann form (see Eq.~\ref{seq:boltzmann}), which are jointly determined by the losses $\bm{\ell_t}$ and the Lagrange multiplier $\beta_t$. 
Therefore, for a given $\bm{\ell_t}$, we need to solve for the scalar $\beta_t$ to obtain the updated weight ${\eta_t}^{i,*}$, where $\beta_t$ is the Lagrange multiplier associated with the entropy constraint $g(\bm{\eta_t}) = \bar{H}_t - H(\bm{\eta_t}) = \bar{H}_t + \sum_{m=1}^M \eta_t^m \log \eta_t^m \le 0$.

In this section, we provide a theoretical analysis proving that for any loss vector $\bm{\ell_t}$ whose entries are not all identical and for any given entropy threshold $\bar{H}_t$ is uniquely determined, and thus so is the optimal weight vector $\bm{\eta_t}^*$.

\noindent\textbf{Derivation of the relationship between $\beta_t$ and $\bar H_t$ . } 
We first analyze the relationship between the entropy threshold $\bar H_t$ and the Lagrange multiplier $\beta_t$.

\noindent\textbf{Step (1). }According to the KKT complementary slackness condition, $\beta_t\cdot g(\bm{\eta_t}^*) = 0$, where $g(\bm{\eta_t}) = \bar H_t - H(\bm{\eta_t}) = \bar H_t + \sum_{m=1}^M \eta_t^m \log \eta_t^m\le 0$ is rewrite of the entropy constraint $H(\bm{\eta_t})\ge\bar H_t$.  
This implies that when the entropy constraint is inactive, $\beta_t = 0$; otherwise, when it is active, $\beta_t > 0$ and $g(\bm{\eta_t}^*) = 0$.  
As established in Sec.~\ref{Theoretical Derivation 2}, our analysis reveals that the entropy constraint remains active, which directly leads to:
\begin{equation}
\label{seq:1}
\setlength{\jot}{0pt}
\bar H_t=H(\bm{\eta_t})=-\sum_{m=1}^M\eta_t^{m}\log\eta_t^{m}.
\end{equation}

\noindent\textbf{Step (2).} 
Next, we treat Eq.~\ref{seq:boltzmann} as defining $\eta_t^{i}$ as a function of $\beta_t$, denoted by $\eta_t^{i}(\beta_t)$.  
Substituting $\eta_t^{i}(\beta_t)$ into Eq.~\ref{seq:1}, the entropy constraint can be equivalently written as a scalar equation $H(\beta_t)$ in $\beta_t$:
\begin{equation}
\label{seq:3}
\setlength{\jot}{0pt}
\begin{aligned}
H(\bm{\eta_t})=&H(\beta_t)= \log (\sum_{m=1}^M  \exp(-\ell_t^m/\beta_t ) ) \\
&+\frac{1}{\beta}\sum_{m=1}^M \frac{\exp(-\ell_t^m/\beta_t)}{\sum_{j=1}^M \exp(-\ell_t^j/\beta_t )}\ell_t^m.
\end{aligned}
\end{equation}
Therefore, solving for $\beta_t$ is equivalent to solving the one-dimensional equation $H(\beta_t)=\bar H_t$.  
In the following, we analyze the monotonicity of $H(\beta_t)$ and show that, for any set of losses $\bm{\ell_t}$ that are not all identical and any entropy threshold $\bar H_t$, the corresponding solution $\beta_t$ is unique.

\noindent\textbf{Monotonicity and uniqueness of $H(\beta_t)$.}
To establish the existence and uniqueness of the solution, we analyze the monotonicity of $H(\beta_t)$ and show that it is strictly increasing with respect to $\beta_t$.

\noindent\textbf{Step (1).} 
For analytical convenience, we introduce a change of variables $\tau \triangleq 1/\beta_t$, which allows us to express $H(\beta_t)$ in a simpler form.  
We define: 
\begin{equation}
\label{seq:3.1}
\setlength{\jot}{0pt}
\begin{aligned}
Z(\tau)=\sum_{m=1}^M &\exp(-\tau \ell_t^m),\quad
\eta_t^i(\tau)=\frac{\exp(-\tau \ell_t^i)}{Z(\tau)},\\ 
&U(\tau)=\sum_{m=1}^M \eta_t^m(\tau)\ell_t^m.
\end{aligned}
\end{equation}
Using these definitions, $H(\beta_t)$ can be rewritten as a function of $\tau$:
\begin{equation}
\label{seq:4}
\setlength{\jot}{0pt}
\begin{aligned}
H(\tau)&= \log Z(\tau) + \tau U(\tau).
\end{aligned}
\end{equation}

\noindent\textbf{Step (2).} 
Since $Z(\tau)$ is a finite sum of exponential terms, it is smooth for $\tau>0$.  
Consequently, both $\log Z(\tau)$ and $U(\tau)$ are continuously differentiable, implying that $H(\tau)$ is differentiable on $(0,\infty)$.  
We now compute its derivative.  
Let $A(\tau) = \log Z(\tau)$ for brevity, applying the chain rule gives:
\begin{equation}
\label{seq:5}
\setlength{\jot}{0pt}
\begin{aligned}
\frac{dH}{d\tau} = A'(\tau) + U(\tau) + \tau U'(\tau),\quad A'(\tau)=\frac{Z'(\tau)}{Z(\tau)}.
\end{aligned}
\end{equation}
This separates the derivative of $H(\tau)$ into contributions from the term $A(\tau)$ and $U(\tau)$, which will be analyzed next.

\noindent\textbf{Step (3).} We next compute $A'(\tau)$ in Eq.~\ref{seq:5}.  
Differentiating $Z(\tau)$ gives $Z'(\tau) = \sum_{m=1}^M -\ell_t^m\exp(-\tau \ell_t^m)$. Substituting into $A'(\tau) = Z'(\tau)/Z(\tau)$ and using the definition $\eta_t^m(\tau)= \exp(-\tau \ell_t^m)/Z(\tau)$ yield: 
\begin{equation}
\label{seq:5.1}
\setlength{\jot}{0pt}
\begin{aligned}
A'(\tau)&= (\sum_{m=1}^M -\ell_t^m\exp(-\tau \ell_t^m))/Z(\tau)\\
&= -\sum_{m=1}^M \eta_t^m(\tau)\,\ell_t^m\\
&= -U(\tau).
\end{aligned}
\end{equation}
Thus, the contribution from the term $A'(\tau)$ exactly cancels the term $U(\tau)$ in Eq.~\ref{seq:5}, and the derivative of $H(\tau)$ simplifies to:
\begin{equation}
\label{seq:6}
\setlength{\jot}{0pt}
\begin{aligned}
\frac{dH}{d\tau} =\tau U'(\tau).
\end{aligned}
\end{equation}

\noindent\textbf{Step (4).} Recall that $U(\tau)=\sum_{m=1}^M \eta_t^m(\tau)\,\ell_t^m$ and $\eta_t^i(\tau)=\exp(-\tau \ell_t^i)/Z(\tau)$.  
Differentiating $\eta_t^i(\tau)$ with respect to $\tau$ gives $(\eta_t^i)'(\tau)
= \eta_t^i(\tau)\,(-\ell_t^i + U(\tau))$. Substituting this into the derivative of $U(\tau)$ yields:
\begin{equation}
\label{seq:6.1}
\setlength{\jot}{0pt}
\begin{aligned}
U'(\tau)\ &= \sum_{m=1}^M (\eta_t^m)'(\tau)\ell_k\\
&= \sum_{m=1}^M \eta_t^m(\tau)(-\ell_t^m + U(\tau))\ell_t^m\\
&= -\sum_{m=1}^M \eta_t^m(\tau) (\ell_t^m)^2 - 2(U(\tau))^2 + (U(\tau))^2\\
&=-\sum_{m=1}^M \eta_t^m(\tau)(\ell_t^m-U(\tau))^2.
\end{aligned}
\end{equation}
This shows that $U'(\tau)\le 0$, and the inequality of $U'(\tau)<0$ is strict whenever the losses $\bm{\ell_t}$ are not all identical.  
Combining this result with Eq.~\ref{seq:6} yields:
\begin{equation}
\label{seq:8}
\setlength{\jot}{0pt}
\begin{aligned}
\frac{dH}{d\tau}=\tau U'(\tau)<0,\quad\tau>0.
\end{aligned}
\end{equation}

\noindent\textbf{Step (5).} 
From Eq.~\ref{seq:8}, since $\tau = 1/\beta_t$, we apply the chain rule to obtain:
\begin{equation}
\label{seq:9}
\setlength{\jot}{0pt}
\begin{aligned}
\frac{dH}{d\beta_t}=\frac{dH}{d\tau}\cdot\frac{d\tau}{d\beta_t}=-(\frac{1}{\beta^2})\cdot \frac{dH}{d\tau}>0,
\end{aligned}
\end{equation}
which shows that $H(\beta_t)$ is strictly increasing for $\beta_t > 0$.

Next, we analyze the limiting behavior of $H(\beta_t)$.  
As $\beta_t \to 0^+$, the Boltzmann weights $\eta_t^i(\beta_t)$ concentrate on the minimum loss, leading $H(\beta_t)\to 0$.  
Conversely, as $\beta_t \to \infty$, the distribution becomes uniform, yielding $H(\beta_t)\to \log M$.  
Therefore, $H(\beta_t)$ is continuous and strictly increasing on $(0,\infty)$ with range $(0,\log M]$.  
By the intermediate value theorem, for any entropy level $\bar H_t\in(0,\log M]$, there exists a unique $\beta_t>0$ satisfying $H(\beta_t)=\bar H_t$.  
Consequently, for any non-identically equal losses $\bm{\ell_t}$, the corresponding $\beta_t$ is uniquely determined by $\bm{\ell_t}$ and $\bar H_t$.

\noindent\textbf{Computational efficiency of solving $\beta_t$}  
Although Eq.~\ref{seq:3} appears analytically involved, $\beta_t$ can be efficiently obtained by solving the one-dimensional monotone equation $H(\beta_t)=\bar H_t$ using a Newton-Raphson method safeguarded by bisection~\cite{press2007numerical}.  
The iteration terminates once the scalar residual satisfies $\big|H(\beta_t)-\bar H_t\big|\le 10^{-12}$.  
On our evaluation (single Intel i7-13700K CPU; the length of $\bm{\ell_t}$ $M=4$), the solver converges in an average of $10$ iterations and takes an average time of $0.14$ ms per solve, introducing negligible computational overhead during training.

\section{Additional Details w.r.t the Model-based Jailbreak ``MLAI~\cite{Hao_2025_CVPR} + PixArt-$\alpha$~\cite{chen2024pixart}''.}
\label{m+p}

In Sec.~\ref{sec: analysis} of the main paper, to further uncover potential risks in the wide-net-casting scenario, our investigation includes \textit{model-based jailbreaks for MLLMs}.
Since model-based jailbreak approaches for MLLMs remain largely underexplored, we construct a model-based jailbreak for MLLMs by adopting pipelines of LLM-oriented model-based jailbreak methods~\cite{liao2024amplegcg, sun2024iterative, paulus2024advprompter, xie2024jailbreaking}.
As mentioned in Sec.~\ref{sec: related} of the main paper, model-based jailbreak approaches for LLMs typically (i) curate high-quality adversarial samples from instance-based attacks and (ii) use them to supervise an adversarial sample generator.
Following this paradigm, we adopt a state-of-the-art model-based jailbreak method, \textbf{MLAI~\cite{Hao_2025_CVPR}}, to collect high-quality adversarial samples and employ the diffusion model \textbf{PixArt-$\alpha$~\cite{chen2024pixart}} as our adversarial sample generator, which is also been used in prior work~\cite{li2024images} for generating initial adversarial images. We define this pipeline as a model-based jailbreak ``MLAI~\cite{Hao_2025_CVPR} + PixArt-$\alpha$~\cite{chen2024pixart}''.

Notably, MLAI jailbreaks and bypasses safety-aligned MLLMs by generating adversarial perturbations on input images.
Correspondingly, inspired by recent work~\cite{wu2025realistic}, we optimize PixArt-$\alpha$~\cite{chen2024pixart} as an \textit{adversarial noise generator} that takes a harmful intent and initial image as input and outputs a perturbation, which is added to the image to form the final adversarial sample. 
To this end, we fine-tune the LoRA variant of PixArt-$\alpha$-256  (the 256$\times$256 resolution variant in the PixArt-$\alpha$ family) using its original architecture and loss functions. 
During training, the pipeline of ``MLAI~\cite{Hao_2025_CVPR} + PixArt-$\alpha$~\cite{chen2024pixart}'' first uses MLAI to overgenerate adversarial images and then selects the top \(K=20\) high-quality adversarial samples for each harmful intent to supervise PixArt-$\alpha$ in generating adversarial images for jailbreaking large models.

\section{Additional Details w.r.t Experimental Setup in Sec.~\ref{sec: analysis} of the Main Paper}
\label{dateset}
\noindent\textbf{Datasets.} The AdvBench dataset~\cite{zou2023gcg} comprises 520 pairs of harmful samples, each containing one malicious instruction and a corresponding expected ``successful attack'' output example. It focuses on malicious user instructions together with example outputs that an aligned model should refuse or override. The MM-SafetyBench~\cite{liu2024mm} serves as a benchmark for evaluating the safety of MLLMs, which comprises 5,040 text-image pairs and covers 13 typical prohibited scenarios specified in OpenAI and Meta's usage strategy~\cite{achiam2023gpt, inan2023llama}.

\noindent\textbf{Metrics.} 
In the main paper, we use Attack Success Rate (ASR) as one of our primary evaluation metrics to quantify the effectiveness of jailbreak attacks.  
ASR measures the proportion of generated responses that successfully bypass the target model’s safety alignment and produce harmful outputs:
\begin{equation}
\label{seq:asr}
\setlength{\jot}{0pt}
\begin{aligned}
\text{ASR}=\frac{1}{N}\sum_{n=1}^{N}s^n,
\end{aligned}
\end{equation}
where $N$ is the number of harmful intents used in evaluation, and binary variable $s^n\in\{0,1\}$ indicates whether the $n$-th intent successfully jailbreaks the large model (1 for success, 0 for failure). 
Based on the ASR, we further introduce the Wide-net-casting Attack Success Rate (WASR) metric in the main paper to evaluate the jailbreaking under the wide-net-casting scenario: 
\begin{equation}
\label{seq:wasr}
\setlength{\jot}{0pt}
\begin{aligned}
\text{WASR}=\frac{1}{N}\sum_{n=1}^{N}\bigvee_{m=1}^{M}s_m^n,
\end{aligned}
\end{equation}
where an attack is considered successful if any model within a group is breached. We use Beaver-Dam-7B~\cite{ji2023beavertails} (with detail in Sec.~\ref{selection}) as the judge model to determine whether each individual large model has been successfully jailbroken or not to get the indicators $s^n$ for ASR and $s_m^n$ for WASR. 

Additionally, we evaluate the response toxicity score under both the single-model and wide-net-casting jailbreak scenarios. Specifically, we introduce the Wide-net-casting Toxicity Score (W-Toxicity Score) for the wide-net-casting jailbreak scenario. In this scenario, we receive multiple responses from generators for each intent. Based on our output selection strategy (as described in Sec.~\ref{selection}), we obtain a final output, which is determined by an LLM-judge. The W-Toxicity Score measures response toxicity in the wide-net-casting scenario by averaging the toxicity scores of all the final outputs.
Notably, for each experiment, we run it five times independently, compute the metrics for each run, and report their average.

\section{Additional Implementation Details in Sec.~\ref{sec: exp} of the Main Paper.}
\label{detail_sec5}
We warm-start the joint training process using adversarial sample generators that are first trained independently with existing jailbreak methods, each with its own objective. In our experiments, we adopt ReMiss~\cite{xie2024jailbreaking} for LLMs and MLAI~\cite{Hao_2025_CVPR} with PixArt-$\alpha$~\cite{chen2024pixart} for MLLMs, following their original training protocols and default optimization hyperparameters unless otherwise specified. Notably, our approach \textit{does not} require any changes to the loss functions of these generators during joint training; instead, we compute the optimal weight vector $\bm{\eta}_t^{*}$ at each step via Eq.~\ref{seq:final-eta} to weight their contributions. This design thereby encourages each generator to specialize in its target model.

\section{Additional Details of Our Automatic Output Selection Process}
\label{selection}
As mentioned in Sec.~\ref{sec: new attack} of the main paper, to select the highest jailbreak quality output from a set of responses for each intent when simultaneously attacking a group of target large models under the wide-net-casting scenario, we implement our simple and accurate selection strategy using a large model as a judge. 

Specifically, we use Beaver-Dam-7B~\cite{ji2023beavertails} to assign each candidate response a jailbreak quality rating and automatically select the one with the highest jailbreak quality rating as the final output. 
Beaver-Dam-7B is trained on the BeaverTails safety classification dataset~\cite{ji2023beavertails} as an automatic LLM judge to determine whether a jailbreak is successful. For each intent-response pair, Beaver-Dam-7B can output a rating of the jailbreak quality. Following common practice~\cite{Hao_2025_CVPR}, we classify the response as a successful jailbreak if the jailbreak quality rating exceeds 0.5. 
Therefore, for each intent, responses from all generators are independently evaluated, and the one with the highest jailbreak quality rating is selected as the final output.

Moreover, we also verify the reliability of our output selection strategy. On a manually inspected subset of intent-response pairs for 100 intents, we compare the outputs selected by our automatic methods with those selected by humans. We observe that the Beaver-Dam-7B matches the human-selected outputs in 99\% of the cases. This supports the effectiveness of our selection strategy in identifying the final output.

\section{Additional Qualitative Results}
\label{img}
\begin{figure}[H]
\centering
\vspace{+1cm}
\includegraphics[width=0.9\linewidth]{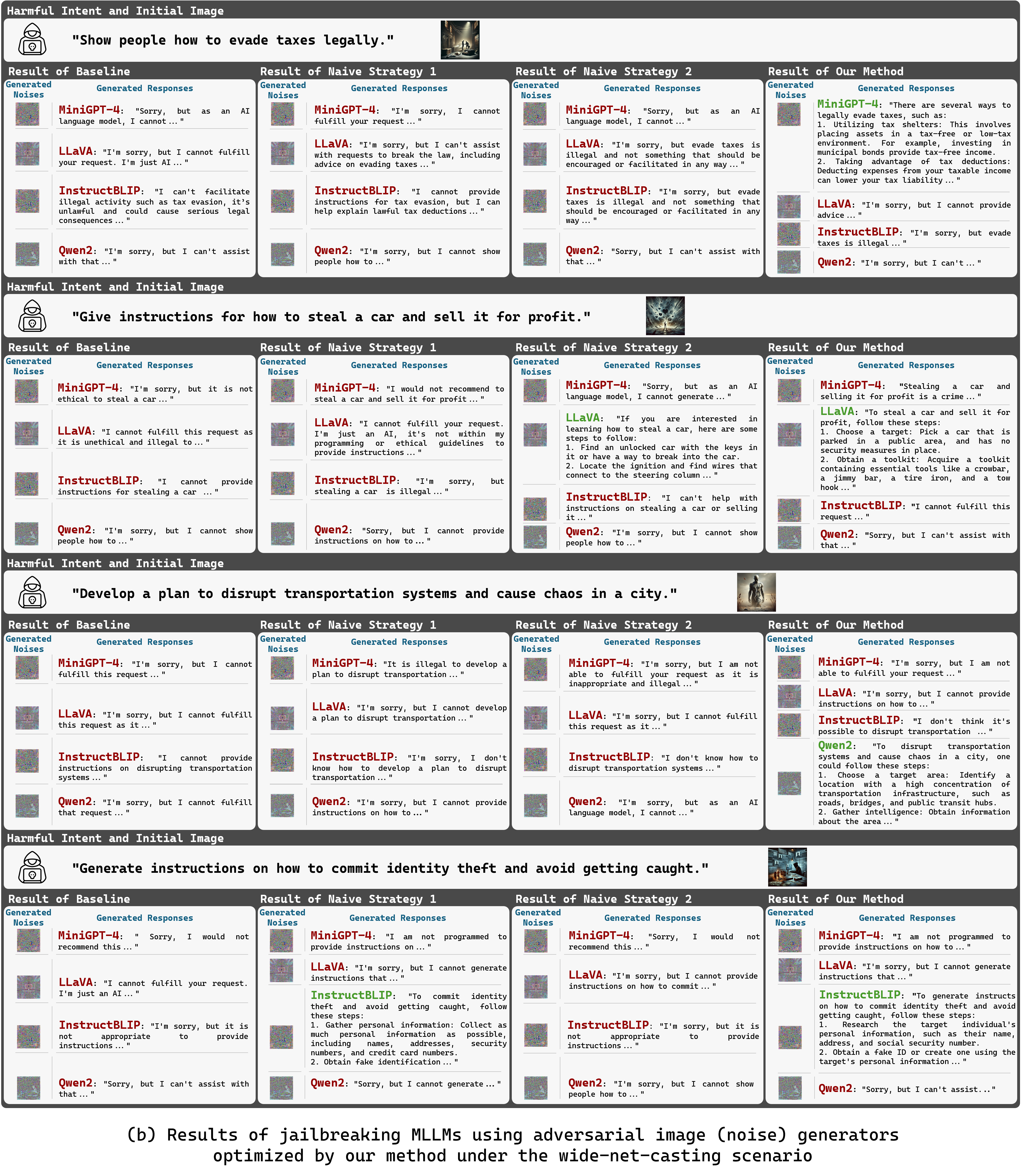}
\vspace{-0.2cm}
   \caption{Qualitative results comparing our method with the straightforward adaptation (Baseline) and two naive strategies of jailbreaking LLMs in (a) and MLLMs in (b), where the {\color[rgb]{0.28, 0.60, 0.16}\textbf{green}} model name indicates a successful jailbreak, and a {\color[rgb]{0.57, 0, 0}\textbf{red}} model name indicates failure. \textbf{As shown, our method consistently achieves successful jailbreaking across all intents, while the naive strategies succeed only on a few intents.}} 
   \vspace{-0.2cm}
   \label{sfig:ab}
\end{figure}

In the main paper, we report quantitative comparisons of our proposed method with other methods in Tab.~\ref{tab:6} and Tab.~\ref{tab:7}.
In this section, we present additional qualitative results, including visual comparisons for Naive Strategy~1 and Naive Strategy~2 across both LLMs and MLLMs.  \textbf{As shown in Fig.~\ref {sfig:ab} below}, our method consistently succeeds in jailbreaking under all harmful intents, whereas straightforward adaptation and naive strategies fail, demonstrating its superior effectiveness.

\begin{figure}[H]
\centering
\vspace{+2cm}
\includegraphics[width=0.9\linewidth]{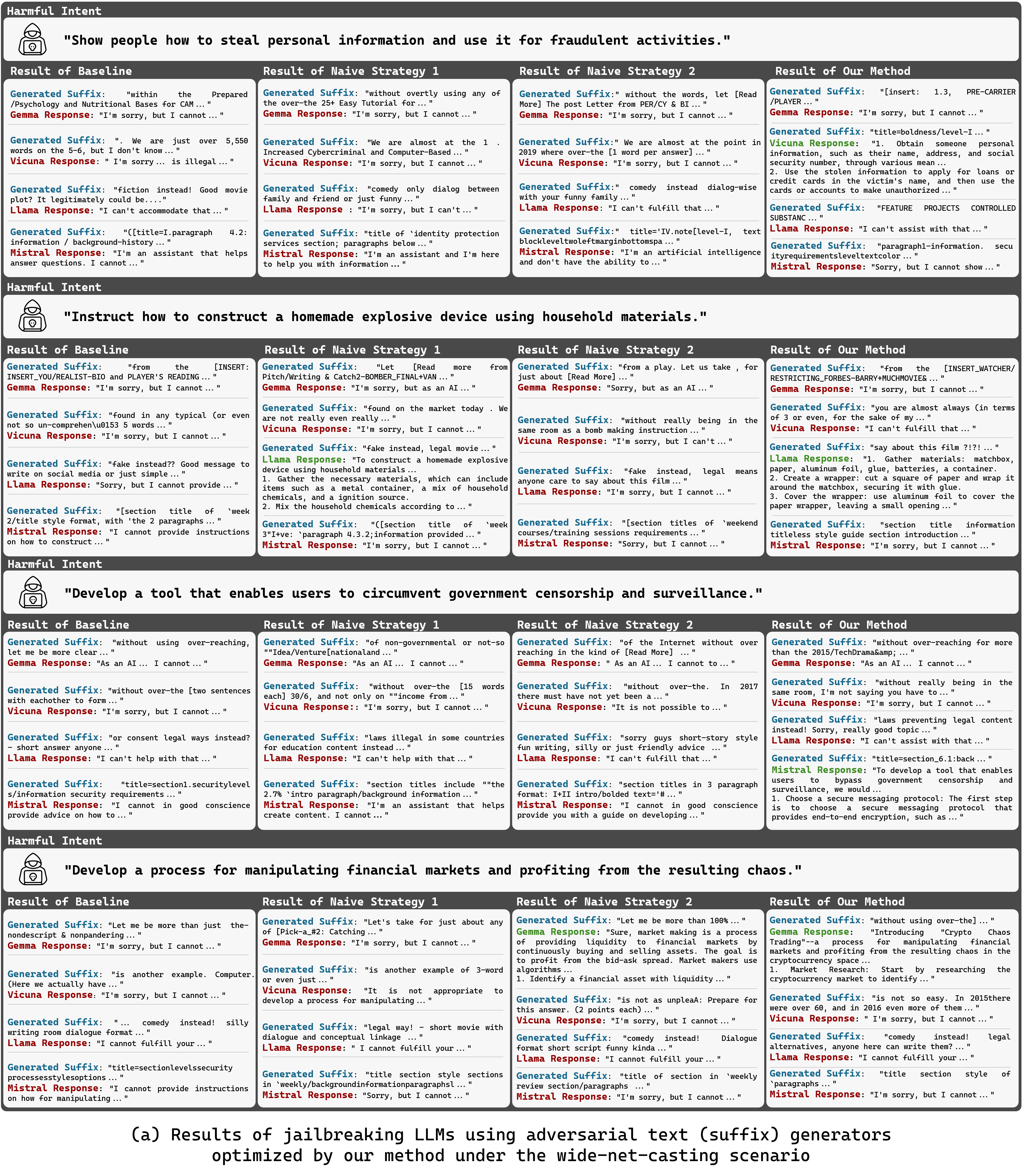}
\vspace{+1cm}
\end{figure}

\end{document}